\documentclass[12pt]{article}
\usepackage{amssymb,graphicx,amsmath,array,verbatim}

\makeatletter
\@addtoreset{equation}{section}

\makeatother

\newcommand {\beq}{\begin{eqnarray}}
\newcommand {\eeq}{\end{eqnarray}}
\def\p{\partial}

\newcommand{\hs}[1]{\hspace{#1 mm}}
\newcommand{\bpm}{\begin{pmatrix}}
\newcommand{\epm}{\end{pmatrix}}

\newcommand{\C}{\mathbb{C}}

\newcommand{\tr}{{\rm Tr}}
\newcommand{\D}{\mathcal D}

\newcommand{\ba}{\left( \begin{array}}
\newcommand{\ea}{\end{array} \right)}
\newcommand{\be}{\begin{equation}}
\newcommand{\ee}{\end{equation}}
\newcommand{\bea}{\begin{eqnarray}}
\newcommand{\eea}{\end{eqnarray}}
\newcommand{\beann}{\begin{eqnarray*}}
\newcommand{\eeann}{\end{eqnarray*}}

\newcommand{\Z}{\mathbb{Z}}

\newcommand{\del}{\partial}

\title{The Moduli Space Metric for\\ 
Well-Separated 
Non-Abelian Vortices}
\author{Toshiaki Fujimori${}^{1}$, Giacomo Marmorini${}^{2}$, Muneto Nitta${}^{2}$,\\ Keisuke Ohashi${}^{3}$ and
Norisuke Sakai${}^{4}$\footnote{Email addresses: 
fujimori(at)th.phys.titech.ac.jp, giacomo(at)phys-h.keio.ac.jp,
nitta(at)phys-h.keio.ac.jp, ohashi(at)gauge.scphys.kyoto-u.ac.jp, 
sakai(at)lab.twcu.ac.jp}
\\
\\
{\it\small 
$^1$ Department of Physics, Tokyo Institute of
Technology, Tokyo 152-8551, Japan}\\
{\it\small
$^2$ Department of Physics, and Research and Education Center for Natural Sciences,}\\  
{\it\small 
Keio University, Hiyoshi 4-1-1, Yokohama, Kanagawa 223-8521, Japan}\\
{\it\small 
$^3$ Department of Physics, Kyoto University, Kyoto 
606-8502, Japan}\\
{\it\small 
$^4$ 
Department of Mathematics, Tokyo Woman's Christian University, 
Tokyo 167-8585, Japan
}\\
}
\begin{document}
\maketitle
\begin{abstract}
The moduli space metric and its K\"ahler potential 
for {\it well-separated} non-Abelian vortices are obtained 
in $U(N)$ gauge theories with $N$ 
Higgs fields in the fundamental representation. 
\end{abstract}

\newpage

\section{Introduction}\label{sec:1}

Solitons, namely smooth localized solutions of nonlinear partial differential equations, have a long history in mathematical and physical
sciences and can now be considered as a subject on their own. The number of physically relevant applications of soliton theory is huge
and ranges from nonlinear optics to astrophysics. While in the mathematical literature the term soliton is mostly associated to integrable
systems, in the framework of modern Lorentz-invariant field theories it refers to smooth localized solutions of field equations, that in
general do not exhibit integrability. A particularly interesting class of solitons is represented by those solutions of the field equations
which satisfy Bogomolnyi-Prasad-Sommerfield (BPS) bound, that is a lower bound for the energy functional. They are topologically stable and
can be shown to actually satisfy first order, instead of second order, partial differential equations that do not involve time derivatives
(BPS equation). Abrikosov-Nielsen-Olesen (ANO) vortices 
\cite{Abrikosov:1956sx} at critical coupling 
, `t Hooft-Polyakov monopoles with massless Higgs field 
\cite{'tHooft:1974qc} 
and instantons in Euclidean
Yang-Mills theory \cite{Belavin:1975fg} 
are prominent examples (a standard reference is \cite{Manton:2004tk}). 
One characteristic feature of BPS solitons is that 
there exist no static forces among them. 
Therefore a large number of soliton configurations 
are allowed with degenerate energy, 
and consequently generic solutions contain moduli parameters 
(collective coordinates). 
The space of solutions of a given set of BPS
equations is called moduli space and is parameterized by those moduli parameters.

The complete characterization of soliton moduli space is not only mathematically attractive, but has deep physical implications. In fact,
while the dynamics of solitons in the full field theory is usually inaccessible (sometimes even numerically), following the idea of Manton
\cite{Manton:1981mp} one can argue that at sufficiently low energies the time evolution is constrained by potential energy to keep the
field configuration close to the moduli space, which is in general finite dimensional. The problem is then reduced to analyze the motion on
the moduli space, which is actually a geodesic motion of the metric induced by the kinetic term of the field theory Lagrangian. 

However, although a number of mathematical structures have been found in various cases, the explicit determination of moduli space metric
can be very difficult in practice. For example, the moduli space of $k$ $SU(2)$ monopoles reveals a hyper-K\"ahler structure, which is
quite restrictive, but the metric is explicitly known only in the $k=2$ case, namely the Atiyah-Hitchin metric \cite{Atiyah:1985dv,Atiyah:1988jp}. Even
in this case the geodesic motion is not integrable, except in a situation where one is allowed to use the asymptotic form of the metric.
The asymptotic metric for well-separated BPS monopoles 
was constructed by Gibbons and Manton \cite{Gibbons:1995yw}.
For other gauge groups the Weinberg-Lee-Yi metric is well known 
\cite{Lee:1996kz}.

For BPS Abelian  vortices (ANO vortices) in flat space 
$\mathbb{R}^2=\mathbb{C}$, 
the $k$-vortex moduli space was shown to be K\"ahler and 
a symmetric product 
${\cal M}_{k} \simeq \mathbb{C}^k/{\cal S}_k$,
where ${\cal S}_k$ denotes symmetrization \cite{Taubes:1979tm}. 
For the metric on it, 
a major step was made in the work of Samols
\cite{Samols:1991ne}, where
a general formula for the metric was given in terms of
local data of the solutions of BPS equations. A K\"ahler potential for such metric could then be found easily (for a direct approach
to the calculation of the K\"ahler potential with different arguments see \cite{Chen:2004xu}). Subsequently Manton and Speight
\cite{Manton:2002wb} calculated the local data for well-separated vortices and, making use of Samols' formula, explicitly wrote down the
asymptotic expression of the moduli space metric for $k$ vortices.
Recently the moduli space metric was given for vortices 
on a hyperbolic space \cite{Krusch:2009tn} in which case 
the system is integrable \cite{Witten:1976ck}.

BPS non-Abelian vortices in more general Higgs models with non-Abelian gauge symmetry were introduced in
\cite{Hanany:2003hp,Auzzi:2003fs} 
(for a review see \cite{review,Eto:2006pg}). 
Such configurations are parametrized not only by position moduli, but also by orientational moduli, that
appear due to the presence of a non-trivial internal color-flavor space; 
it was found that a single vortex moduli space is
\beq 
 {\cal M}_{k=1,N} \simeq {\mathbb C} \times {\mathbb C}P^{N-1} 
\label{eq:single-moduli}
\eeq
for $U(N)$ gauge theory with $N$ Higgs fields in 
the fundamental representation.
The K\"ahler class on ${\mathbb C}P^{N-1}$ was determined 
to be $4\pi/g^2$, with  $g$  the gauge coupling constant
\cite{Shifman:2004dr}.
The analysis of the moduli space 
has gone through many developments especially 
after the introduction of the moduli matrix formalism 
\cite{Isozumi:2004vg}--\cite{Eto:2009wq} 
(for a review of the method see
\cite{Eto:2006pg}). 
The moduli matrix is a matrix whose components are 
holomorphic polynomials of $z$ 
(codimensions of vortices), 
and it contains all moduli parameters in 
coefficients \cite{Isozumi:2004vg}. 
The moduli space of multiple vortices at arbitrary positions 
with arbitrary orientations in the internal space 
was constructed in \cite{Eto:2005yh}.
A general formula for the K\"ahler potential 
on the moduli space 
was obtained in \cite{Eto:2006uw}.  
For separated (not necessary well-separated) 
non-Abelian vortices, 
the moduli space can be written as the symmetric product of 
$k$ copies of the single vortex moduli space (\ref{eq:single-moduli}) 
\cite{Eto:2005yh}:
\beq 
 {\cal M}_{k,N} \leftarrow
({\mathbb C} \times {\mathbb C}P^{N-1})^k/{\cal S}_k .
\label{eq:k-moduli}
\eeq
where the arrow denotes the resolution of sigularities;
The space on the right hand side contains orbifold singularities 
which correspond to coincident vortices, while 
the full moduli space ${\cal M}_{k,N}$ on the left hand side 
should be regular. 
By evaluating the K\"ahler potential 
\cite{Eto:2006uw} of the moduli space at linear order,
it was explicitly shown in \cite{Eto:2006db} 
that the metric is actually regular everywhere even 
at coincident limits of two vortices \cite{Eto:2006cx}.
The head-on-collision of two vortices was also studied  
in \cite{Eto:2006db}.

The purpose of the present paper is to give  
the metric and its K\"ahler potential on the moduli space (\ref{eq:k-moduli}) 
for {\it well-separated} non-Abelian vortices. 
Our main results are the generalization of Samols' formula to the non-Abelian case and, starting from that and
from the asymptotics of non-Abelian vortex solutions \cite{Eto:2009wq}, the derivation of the explicit metric and 
its K\"ahler potential.
The final form of the metric exhibits an evident interplay between spatial (position) and orientational moduli, opening up a rich variety of
possibly interesting dynamics, that will be the object of a further investigation \cite{future}.
In this paper we concentrate on local vortices, 
namely vortices in $U(N)$ gauge theories 
with  Higgs fields in the fundamental representation 
in the same number  as 
the number $N$ of colors (while semi-local vortices \cite{Vachaspati:1991dz} 
exist in theories with more fundamental Higgs fields 
\cite{Shifman:2006kd,Eto:2007yv}). 
We also restrict ourselves to $U(N)$ gauge group although 
non-Abelian vortices with gauge group $G \times U(1)$ 
with arbitrary simple group $G$ have been recently constructed in 
\cite{Eto:2008yi}--\cite{Gudnason:2010jq}.
We leave generalizations to those cases as future works.

The paper is organized as follows. In Section~\ref{sec:2} we define the model and review the construction of non-Abelian vortices in the
moduli matrix formalism. In Section~\ref{sec:3} we find the non-Abelian extension of the Samols' formula for the metric on the moduli space and
in Section~\ref{sec:4} we show how it can be made explicit 
in the case of well-separated vortices, which 
means to find the asymptotic
metric and its K\"ahler potential. In Section~\ref{sec:5} we obtain the latter result by means of a more physical method, namely point-particle approximation. Some
details of the calculation are given in Appendix~\ref{appendix:DE}.

\section{Review of non-Abelian local vortices}\label{sec:2}
\subsection{Lagrangian and BPS equations}\label{subsec:Lagrangian}
Let us consider a $U(N)$ gauge theory in $(2+1)$-dimensional spacetime with gauge fields $w_\mu$ for $U(1)$, $W_\mu^a~(a=1,\ldots,N^2-1)$
for $SU(N)_C$ and $N$ Higgs fields $H^A~(A=1,\ldots,N)$ in the fundamental representation of the $SU(N)_C$ gauge group. The Lagrangian of
the
theory takes the form 
\beq
\mathcal L &=& - \frac{1}{4e^2} (f_{\mu \nu})^2 - \frac{1}{4g^2} (F^a_{\mu \nu})^2 + (\D^\mu H^A)^\dagger \D_\mu H^A - V, \label{eq:L} \\
V &=& \frac{e^2}{2} ( H^\dagger_A t^0 H^A - \xi )^2 + \frac{g^2}{2} (H^\dagger_A t^a H^A)^2, 
\eeq
where $\xi$ is the Fayet-Iliopoulos parameter, $e$ and $g$ 
are gauge coupling constants for $U(1)$ and $SU(N)_C$, 
respectively. 
Our notation is 
$\D_\mu H^A = (\p_\mu + i w_\mu t^0 + i W_\mu^a t^a) H^A$ 
and $f_{\mu \nu} t^0 + F_{\mu \nu}^a t^a = -i [\D_\mu, \D_\nu]$. 
The matrices
$t^0$ and $t^a$ are the generators of $U(1)$ and $SU(N)$, normalized as
\beq
t^0 = \frac{1}{\sqrt{2N}} \mathbf 1_N, \hs{10} \tr ( t^a t^b ) = \frac{1}{2} \delta^{ab}.
\eeq
As is well known, the Lagrangian Eq.\,\eqref{eq:L} can be embedded into a supersymmetric theory with eight supercharges. The Higgs fields
can also be expressed as an $N$-by-$N$ matrix on which the $SU(N)_C$ gauge transformations act from the left and the $SU(N)_F$ flavor
symmetry acts from the right
\beq
H \rightarrow U_C H U_F^\dagger, \hs{10} U_C \in SU(N)_C, \hs{5} U_F \in SU(N)_F.
\eeq
Using this matrix notation for the Higgs fields, the vacuum condition can be written as
\beq
H H^\dagger - v^2 \, \mathbf 1_N = 0, \hs{10} v^2 \equiv \sqrt{\frac{2}{N}} \xi.\label{eq:vac}
\eeq
The vacuum of this model is in an $SU(N)_{C+F}$ 
color-flavor locking phase, where the vacuum expectation 
values (VEVs) of the Higgs fields are 
\beq
H = v \, \mathbf 1_N.
\eeq
In this vacuum, the mass spectrum is classified according to the representation of $SU(N)_{C+F}$
\beq
m_e = e v, \hs{10} m_g = g v,
\eeq
where $m_e$ is for singlet fields and $m_g$ is for adjoint fields.

Considering a static configuration,  the  BPS bound for the energy reads
\beq
E \ \geq \ - v^2 \int d^2 x \, \tr ( f_{12} t^0 ) \ = \ 2 \pi v^2 k, \hs{10} k \in \Z.
\eeq
The  bound is saturated if the following BPS equations are satisfied: 
\beq
\D_{\bar z} H = 0 , \hs{15} \frac{2}{e^2} f_{12} t^0 + \frac{2}{g^2} F_{12}^a t^a  = H H^\dagger - v^2 \mathbf 1_N, 
\label{eq:BPS} 
\eeq 
where $z=x^1+ix^2$ is a complex coordinate.
To solve the BPS equations, it is convenient to 
rewrite the gauge fields in terms of matrices 
$S_e (\propto \mathbf 1_N) \in \C^\ast$ and $S_g \in SL(N,\C)$ 
\beq
w_{\bar z} t^0 &\equiv& \frac{1}{2} ( w_1 + i w_2)t^0 \,~~~=~ - i S_e^{-1} \bar \p S_e, \label{eq:gauge0} \\ 
W_{\bar z}^a t^a &\equiv& \frac{1}{2} (W_1^a + i W_2^a)t^a ~=~ - i S_g^{-1} \bar \p S_g.
\label{eq:gauge}
\eeq
Then, the first BPS equation $\D_{\bar z} H = 0$ can be solved 
as 
\beq
H = S_e^{-1} S_g^{-1} H_0(z) ,
\label{eq:Higgs}
\eeq
where $H_0(z)$ is an arbitrary $N$-by-$N$ matrix which is holomorphic in $z$.
The second BPS equation becomes \cite{Isozumi:2004vg,Eto:2005yh}
\beq
\frac{4}{m_e^2} \bar \p ( \Omega_e \p \Omega^{-1}_e) + \frac{4}{m_g^2} \bar \p ( \Omega_g \p \Omega_g^{-1}) = v^{-2} H_0 H_0^\dagger \Omega_g^{-1} \Omega_e^{-1} - \mathbf 1_N,
\label{eq:master}
\eeq
where $\Omega_e (\propto \mathbf 1_N)$ and $\Omega_g\in SL(N,\mathbb C)$ are positive-definite Hermitian matrices defined by 
\beq
\Omega_e \equiv S_e S_e^\dagger, \hs{10} \Omega_g \equiv S_g S_g^\dagger.
\label{eq:Omega}
\eeq
We call Eq.\,\eqref{eq:master} the ``master equation'' 
for non-Abelian vortices.

By using the matrices $H_0(z)$, $\Omega_e$ and $\Omega_g$, the BPS equations can be solved by the following procedure. Taking an arbitrary holomorphic
matrix $H_0(z)$, we solve the master equation \eqref{eq:master} in terms of $\Omega_e$ and $\Omega_g$ with the boundary conditions such that
the vacuum equation \eqref{eq:vac} is satisfied at  spatial infinity $|z| \rightarrow \infty$. Explicitly, they are given by
\beq
\Omega_e \rightarrow v^{-2} | \det H_0 |^{\frac{2}{N}} \mathbf 1_N, \hs{10} \Omega_g \rightarrow | \det H_0 |^{-\frac{2}{N}} H_0 H_0^\dagger. 
\label{eq:BD}
\eeq
From the positive-definite hermitian matrices $\Omega_e = S_e S_e^\dagger$ and $\Omega_g = S_g S_g^\dagger$, the matrices $S_e$ and $S_g$
can be determined uniquely up to $U(1) \times SU(N)_C$ gauge transformation $S_e \rightarrow S_e e^{-i \alpha}$, $S_g \rightarrow S_g
U^{-1}$. Then, the physical fields can be obtained via the relations Eq.\,\eqref{eq:gauge0}, Eq.\,\eqref{eq:gauge} and Eq.\,\eqref{eq:Higgs}. 

As a consequence of the definitions 
\eqref{eq:gauge0}, \eqref{eq:gauge} and \eqref{eq:Higgs}, 
the master equation \eqref{eq:master} possesses a symmetry
under the so-called  ``$V$-transformation" \cite{Isozumi:2004vg,Eto:2005yh}
\beq
S_g S_e \rightarrow V(z) S_g S_e, \hs{10} H_0(z) \rightarrow V(z) H_0(z), 
\label{eq:V-transf}
\eeq
where $V(z) \in GL(N,\C)$ is an arbitrary non-singular matrix holomorphic in $z$. Since the physical fields $w_\mu,\,W_\mu$ and $H$
are invariant under  $V$-transformations, \eqref{eq:V-transf} defines an equivalence relation on the set of holomorphic matrices $H_0(z)$
\beq
H_0(z) \sim V(z) H_0(z).
\eeq
There exists a one-to-one correspondence 
between the equivalence classes $H_0 \sim V H_0$ 
and points on the moduli space of the BPS vortices
\cite{Eto:2005yh,Isozumi:2004vg,Eto:2006pg,Eto:2006db}. 
In this sense, we call $H_0(z)$ the ``moduli matrix" 
and the parameters contained in $H_0(z)$ are 
identified with the moduli parameters of the BPS configurations. 
For example, the vortex positions for a given moduli matrix $H_0(z)$ 
can be determined as follows. 
Since a part of gauge symmetry is restored inside the vortex core, 
the vortex positions can be defined as those points 
on the complex plane at which the rank of the matrix $H$ 
becomes smaller than $N$, 
namely they can be determined as 
the zeros of the holomorphic polynomial $\det H_0$. 

\subsection{Single vortex configurations}\label{subsec:single}
As an example, let us consider configurations of a single vortex located at $z=z_0$. Since zeros of the polynomial $\det H_0$ correspond to
the vortex position, we consider the set of moduli matrices whose determinant is $\det H_0 = z-z_0$. For example, in the case of $N=2$, any
moduli matrix with $\det H_0 = z - z_0$ is $V$-equivalent to the moduli matrix of the form \cite{Eto:2004rz}
\beq
H_0 = \ba{cc} z - z_0 & 0 \\ - \beta & 1 \ea \sim \ba{cc} 1 & - \tilde \beta \\ 0 & z - z_0 \ea, \hs{10} \tilde \beta = \frac{1}{\beta}.
\eeq
In addition to the translational moduli parameter $z_0$, there exists one parameter $\beta$ that can be viewed as an inhomogeneous
coordinate of $\C P^1$. This internal degree of freedom, which is called the orientation, corresponds to the Nambu-Goldstone zero mode of
$SU(2)_{C+F}$ symmetry broken by the vortex. The homogeneous coordinate of $\C P^1$ can also be extracted from the moduli matrix as follows.
Since ${\rm rank} \, H_0$ drops at $z = z_0$, there exists 
an eigenvector of $H_0$ with the null eigenvalue 
at $z=z_0$. 
In other words, there exists a constant $N$-vector 
$\boldsymbol \phi$ such that
\beq
(H_0 \boldsymbol \phi) \big|_{z=z_0} = 0, \hs{10} \boldsymbol \phi \propto \ba{c} 1 \\ \beta \ea.
\eeq
The vector $\boldsymbol \phi$ is called the orientational vector and corresponds to the homogeneous coordinates of $\C P^1$. 
In the case of general $N$, a single vortex configuration breaks $SU(N)_{C+F}$ down to $SU(N-1) \times U(1)$, so that the orientational
moduli space is 
\beq
\C P^{N-1} = \frac{SU(N)}{SU(N-1) \times U(1)}.
\eeq
In this case, the generic moduli matrix with $\det H_0 = z-z_0$ is equivalent to
\beq
H_0 = \ba{cc} z - z_0 & 0 \\ - \vec \beta & \mathbf 1_{N-1} \ea,
\label{eq:single}
\eeq
where $N-1$ parameters $\vec \beta$ are the inhomogeneous 
coordinates parameterizing the internal orientation 
$\C P^{N-1}$. 
As in the case of $N=2$, the orientational vector can also 
be defined by $(H_0 \boldsymbol \phi)|_{z=z_0} = 0$, 
corresponding to the homogeneous coordinates of $\C P^{N-1}$. 

Next, let us briefly review some properties of the solution of the master equation \eqref{eq:master} for the single vortex configuration \cite{Eto:2009wq}. First, note that the moduli matrix Eq.\,\eqref{eq:single} can be rewritten as
\beq
H_0 ~=~ V(z) \ba{cc} z-z_0 & 0 \\ 0 & \mathbf 1_{N-1} \ea U 
\eeq
where the matrices $V(z) \in SL(N,\C)$ and $U \in SU(N)_F$ are given by
\beq
V &=& \ba{cc} \frac{1}{\sqrt{1+|\beta|^2}} & - \frac{(z-z_0) \vec \beta^\dagger}{\sqrt{1+|\vec \beta|^2}} \\ 0 & \left( \mathbf 1 - \mathbf P \right) + \sqrt{1+|\beta|^2} \mathbf P \ea, \label{eq:V} \\
U &=& \ba{cc} \frac{1}{\sqrt{1+|\beta|^2}} & \frac{\vec \beta^\dagger}{\sqrt{1+|\beta|^2}} \\ - \frac{\vec \beta}{\sqrt{1+|\beta|^2}} & \left( \mathbf 1 - \mathbf P \right) + \frac{1}{\sqrt{1+|\beta|^2}} \mathbf P \ea. \label{eq:U}
\eeq
and the $(N-1)$-by-$(N-1)$ matrix $\mathbf P$ is defined by
\beq
\mathbf P \equiv \frac{\vec \beta \vec \beta^\dagger}{|\vec \beta|^2}.
\eeq
It is convenient to use the following ansatz for $\Omega_e$ and $\Omega_g$
\beq
\Omega_e = v^{-2} e^{\psi_e \mathbf 1_N}, \hs{10} \Omega_g = V e^{\psi_g T} V^\dagger, 
\label{eq:ansatz}
\eeq
where $\psi_e$ and $\psi_g$ are smooth 
real functions and the matrix $T \in su(N)$ is given by
\beq
T \equiv {\rm diag} \, (N-1,\,-1,\,\cdots,\,-1).
\eeq
Then, the master equation \eqref{eq:master} reduces to the following two equations for the functions $\psi_e$ and $\psi_g$ 
\beq
\frac{4N}{m_e^2} \p \bar \p \psi_e &=& e^{\psi_g-\psi_e} \left( 1 - |z-z_0|^2 e^{-N \psi_g} \right) + N (1 - e^{\psi_g - \psi_e} ), \\
\frac{4N}{m_g^2} \p \bar \p \psi_g &=& e^{\psi_g-\psi_e} \left( 1 - |z-z_0|^2 e^{-N \psi_g} \right).
\eeq
The boundary conditions for $\psi_e$ and $\psi_g$ can be read from Eq.\,\eqref{eq:BD} as
\beq
\psi_e \rightarrow \frac{1}{N} \log |z-z_0|^2, \hs{10} \psi_g \rightarrow \frac{1}{N} \log |z-z_0|^2.
\eeq
By using the relation Eq.\,\eqref{eq:Omega} and choosing an appropriate gauge, we obtain the matrices $S_e$ and $S_g$ in terms of the functions $\psi_e$ and $\psi_g$ as
\beq
S_e = v^{-1} e^{\frac{1}{2} \psi_e \mathbf 1_N}, \hs{10} S_g = V e^{\frac{1}{2} \psi_g T } U. 
\label{eq:single_S}
\eeq
Then, the solution can be obtained through Eq.\,\eqref{eq:gauge} and Eq.\,\eqref{eq:Higgs} as
\beq
H \ \, &=& v \, e^{-\frac{1}{2} ( \psi_e \mathbf 1_N + \psi_g T_\beta)} e^{\frac{1}{N} (\mathbf 1_N + T_\beta)\log (z-z_0)},
\label{eq:single_H} \\
w_{\bar z} t^0 &=& - \frac{i}{2} \bar \p \psi_e \mathbf \, 1_N, \label{eq:single_w} \\
W_{\bar z}^a t^a &=& -\frac{i}{2} \bar \p \psi_g \, T_\beta, \label{eq:single_W}
\eeq
where we have defined the matrix $T_\beta$ by
\beq
T_\beta ~\equiv~ U^\dagger T U
~=~ N \frac{{\boldsymbol \phi \boldsymbol\phi}^\dagger}{|\boldsymbol \phi|^2}
-\boldsymbol 1_N, \hs{10}
\boldsymbol \phi \propto \ba{c} 1 \\ \vec \beta \ea.
\label{eq:tbeta}
\eeq 
Now let us look at the asymptotic forms of the single vortex solution. 
The functions $\psi_e$ and $\psi_g$ 
behave near the vortex core as 
\beq 
\psi_e &=& a_e + \frac{m_e^2}{4N} \left[ e^{a_g - a_e} 
+ N ( 1 - e^{a_g - a_e}) \right] |z-z_0|^2 
+ \mathcal O(|z-z_0|^4) , \label{eq:expansion1} \\
\psi_g &=& a_g + \frac{m_g^2}{4N} e^{a_g - a_e} 
|z-z_0|^2 + \mathcal O(|z-z_0|^4) , \label{eq:expansion2}
\eeq
where $a_e$ and $a_g$ are constants. On the other hand, 
the asymptotic forms of the functions\footnote{
The function $\psi_g, \psi_e$ of the $SU(N), U(1)$ parts 
here are related to those in Ref.\cite{Eto:2009wq} by 
$\psi_g^{\rm here}=\frac{1}{N-1}\psi_g^{\cite{Eto:2009wq}}
+\frac{1}{N}\log |z-z_0|^2$, and 
$\psi_e^{\rm here}=\psi_e^{\cite{Eto:2009wq}}+\frac{1}{N}\log |z-z_0|^2$. 
Consequently, the coefficient $c_g^{\rm here}=c_g^{\cite{Eto:2009wq}}/(N-1)$. 
} 
 $\psi_e$ and $\psi_g$ 
for large $|z-z_0|$ are given by \cite{Eto:2009wq} 
\beq
\psi_e &=& \frac{1}{N} \log |z-z_0|^2 + c_e K_0(m_e |z-z_0|) 
+ \mathcal O( e^{- 2 m_e |z-z_0|},\, e^{-2 m_g|z-z_0|} ) , \label{eq:asym1} \\
\psi_g &=& \frac{1}{N} \log |z-z_0|^2 + c_g K_0 ( m_g |z-z_0|) 
+ \mathcal O( e^{- 2 m_g |z-z_0|},\, e^{-(m_e+m_g)|z-z_0|} ) \label{eq:asym2},
\eeq 
where $K_0(m|z-z_0|)$ is the modified Bessel function of the second kind. 
The constants $c_e$ and $c_g$ depend on the ratio $m_g/m_e$ and $N$. 
In  case of $m_e\ge 2m_g$, the term proportional to 
$K_0(m_e|z-z_0|)$ is actually 
not dominant in Eq.(\ref{eq:asym1}) 
compared to the contribution of order $e^{-2m_g|z-z_0|}$.
We then 
concentrate on the case with $m_e<2m_g$ so that the
above form is the proper approximation.  
As we will see, however, Eq.(\ref{eq:asym1}) and Eq(\ref{eq:asym2}) are 
sufficient to determine the metric to the leading order 
even for the case $m_e>2m_g$. 

\section{Effective Lagrangian for non-Abelian local vortices}\label{sec:3}
In this section, we derive a formula for the metric on the moduli space of non-Abelian local vortices which  generalizes the celebrated
Samols' formula for Abelian vortices \cite{Samols:1991ne}.

\subsection{Formula for the metric on the moduli space}\label{formula}
 The moduli space of the BPS vortices is a K\"ahler manifold whose holomorphic coordinates are identified with the complex parameters
contained in $H_0(z)$. The effective low-energy dynamics of the BPS vortices are described by an effective Lagrangian of the form
\beq
L_{\rm eff} = g_{i \bar j} \dot \phi^i \dot{\bar \phi}^j 
 = {\del^2 K \over \del\phi^i \del\bar\phi^j} \dot \phi^i \dot{\bar \phi}^j , 
\eeq
where $\phi^i$ are the holomorphic coordinates, $g_{i \bar j}$ is the metric of the moduli space of non-Abelian vortices, 
and $K$ is the K\"ahler potential.
By using the moduli matrix $H_0(z)$ and 
the solution $(\Omega_e,\Omega_g)$ of the master equation \eqref{eq:master}, 
the K\"ahler potential of the moduli space can be formally\footnote{
To make the K\"ahler potential finite,
we need to add counter terms which can be regarded as 
a K\"ahler transformation. 
} written as \cite{Eto:2006uw}
\beq
K = v^2 \int d^2 x \, \tr\left[ \frac{4}{m_e^2} 
\mathcal K( \Omega_e ) + \frac{4}{m_g^2} 
\mathcal K( \Omega_g ) + \log (\Omega_e \Omega_g) 
+ v^{-2} H_0 H_0^\dagger \Omega_e^{-1} \Omega_g^{-1} \right],
\eeq
where $\mathcal K(\Omega)$ is a quantity\footnote{
The explicit form of $\mathcal K(\Omega)$ is given by
\beq
 \tr \, \mathcal K (\Omega) = \int_0^1 ds \int_0^s d t \, \tr 
\left[ \bar \p \omega e^{t \omega} \p \omega e^{-t \omega} \right], \hs{10} \omega \equiv \log \Omega. \notag  
\eeq
}
 which satisfies the following identity for a small variation $\Omega \rightarrow \Omega + \delta \Omega$
\beq
\delta \left[ \mathcal \int d^2 x \, \tr \, \mathcal K(\Omega) \right] = \int d^2 x \, \tr \left[ \bar \p ( \Omega \p \Omega^{-1} ) \delta \Omega \Omega^{-1} \right]. 
\eeq
As a consequence, by varying the K\"ahler potential $K$ with respect to $\Omega_e$ and $\Omega_g$, we can show that $K$ is minimized by the
solution of the master equation \eqref{eq:master}. Therefore, the derivatives of the K\"ahler potential with respect to the moduli
parameters are given by
\beq
\frac{\p}{\p \phi^i} K = \int d^2 x \, \tr \left( \frac{\p}{\p \phi^i} H_0 H_0^\dagger \Omega_e^{-1} \Omega_g^{-1} \right).
\eeq
Note that $H_0^\dagger$ is anti-holomorphic in $\phi^i$. 
From this property of the K\"ahler potential, we obtain a simple form of the effective Lagrangian 
\beq
L_{\rm eff} ~=~ \frac{\p^2 K}{\p \phi^i \p \bar \phi^j} \dot \phi^i \dot{\bar \phi}^j ~=~ \int d^2 x \, \delta_t^\dagger \tr ( \delta_t H_0 H_0^\dagger \Omega_g^{-1} \Omega_e^{-1} ),
\label{eq:metric}
\eeq
where the differential operators $\delta_t$ and $\delta_t^\dagger$ are defined by
\beq
\delta_t = \dot \phi^i \frac{\p}{\p \phi^i}, \hs{10} \delta_t^\dagger = \dot{\bar \phi}^i \frac{\p}{\p \bar \phi^i}.
\eeq

Now we  rewrite the effective Lagrangian in terms of local data in the neighborhood of each vortex. Let us assume that $\det H_0$ has zeros
at $z=z_I~(I=1,2,\cdots)$, namely
\beq
\det H_0(z) = \prod_{I=1}^k (z-z_I).
\eeq
Let $D_\epsilon^I$ be the disk of radius $\epsilon$ centered at $z=z_I$
\beq
D_\epsilon^I = \big\{ z \in \C ~\big|~ |z-z_I| < \epsilon \big\}.
\eeq
It is convenient to decompose the  domain of integration $\C$ into the   disks $D_\epsilon^I$ and their complement $\C - \bigcup
D_\epsilon^I$. 
Since the integrand in \eqref{eq:metric} does not have any singularity, the integral over the disk $D_\epsilon^I$ vanishes in the zero-radius limit $\epsilon \rightarrow 0$
\beq
\lim_{\epsilon \to 0} \int_{D_\epsilon^I} d^2 x \, \delta_t^\dagger \tr ( \delta_t H_0 H_0^\dagger \Omega_g^{-1} \Omega_e^{-1} ) = 0.
\eeq
Therefore, the effective Lagrangian can be evaluated by integrating over $\C - \sum D_{\epsilon}^I$ and then taking $\epsilon \rightarrow 0$ limit
\beq
L_{\rm eff} = \lim_{\epsilon \rightarrow 0} \int_{\C-\bigcup D_\epsilon^I} d^2x \,\, \delta_t^\dagger \tr ( \delta_t H_0 H_0^\dagger \Omega_g^{-1}
\Omega_e^{-1} ) .
\eeq
Using the master equation \eqref{eq:master} the integrand can be put in the form of a total derivative as
\beq
\delta_t^\dagger \tr ( \delta_t H_0 H_0^\dagger \Omega_g^{-1} \Omega_e^{-1} ) &=& v^2 \, \delta_t^\dagger \tr \left[ \delta_t H_0 H_0^{-1} \left( \frac{4}{m_g^2} \bar \p (\Omega_g \p \Omega_g^{-1}) + \frac{4}{m_e^2} \bar \p (\Omega_e \p \Omega_e^{-1} ) + \mathbf 1_N \right) \right] \notag \\
&=& v^2 \, \bar \p \, \tr \left[ \delta_t H_0 H_0^{-1} \, \delta_t^\dagger \hs{-1} \left( \frac{4}{m_g^2} \Omega_g \p \Omega_g^{-1} + \frac{4}{m_e^2} \Omega_e \p \Omega_e^{-1} \right) \right], \label{eq:integrand}
\eeq
where we used the fact that $\delta_t H_0 H_0^{-1}$ is holomorphic with
respect to $z$ and the moduli parameters $\phi^i$ on 
$\C - \sum D_\epsilon^I$. For  Stokes' theorem, the integral of Eq.\,\eqref{eq:integrand} over $\C - \sum D_\epsilon^I$ can be replaced by an 
integral along the infinitely large circle $S^1_\infty$ and the boundaries of the disks $-\p D_\epsilon^I$. Since the integrand falls off
exponentially at spatial infinity, the contribution from $S^1_\infty$ vanishes and the effective Lagrangian becomes 
\beq
L_{\rm eff} &=& - \frac{v^2}{2i} \sum_{I=1}^k \lim_{\epsilon \rightarrow 0} \int_{\p D_\epsilon^I} dz \, \tr \left[ \delta_t H_0 H_0^{-1} \, \delta_t^\dagger \hs{-1} \left( \frac{4}{m_g^2} \Omega_g \p \Omega_g^{-1} + \frac{4}{m_e^2} \Omega_e \p \Omega_e^{-1} \right) \right].
\eeq
Since each integral picks up the terms which behave as $1/(z-z_I)$, it can be evaluated by expanding the integrand around $z=z_I$. 

First let us consider the case where all the zeros of the polynomial
$\det H_0(z)$ are isolated. In this case, the matrix 
$\delta_t H_0 H_0^{-1}$ has the first order pole at $z=z_I$
\beq
\delta_t H_0 H_0^{-1} = \frac{Q_I}{z-z_I} + \{ \mbox{regular at $z=z_I$} \}.
\eeq
Since the remaining part of the integrand is non-singular, it can be expanded around $z=z_I$ as
\beq
\delta_t^\dagger \hs{-1} \left( \frac{4}{m_g^2} \Omega_g \p \Omega_g^{-1} + \frac{4}{m_e^2} \Omega_e \p \Omega_e^{-1} \right) = \dot{\bar z}_I C_I - \delta_t^\dagger B_I + \mathcal O(|z-z_I|),
\eeq
where the matrices $B_I$ and $C_I$ are defined by
\beq
B_I &\equiv& - \lim_{z \rightarrow z_I} \left[ \frac{4}{m_g^2} \Omega_g \p \Omega_g^{-1} + \frac{4}{m_e^2} \Omega_e \p \Omega_e^{-1} \right] , \\
C_I &\equiv& - \lim_{z \rightarrow z_I} \left[ \frac{4}{m_g^2} \bar \p (\Omega_g \p \Omega_g^{-1}) + \frac{4}{m_e^2} \bar \p ( \Omega_e \p \Omega_e^{-1}) \right].
\eeq
Then, the effective Lagrangian can be written as
\beq
L_{\rm eff} &=& - \pi v^2 \sum_I \tr \left[ Q_I (\dot{\bar z}_I C_I - \delta_t^\dagger B_I) \right]. 
\eeq
From the master equation \eqref{eq:master}, we evaluate $C_I$:
\beq
C_I = \mathbf 1_N - v^{-2} H_0 H_0^\dagger \Omega_g^{-1} \Omega_e^{-1} \big|_{z=z_I}.
\eeq
Therefore, the effective Lagrangian reduces to
\beq
L_{\rm eff} = \pi v^2 \sum_I \Big[ |\dot z_I|^2 + \tr \left( Q_I \delta_t^\dagger B_I \right) \Big],
\label{eq:Leff1}
\eeq
where we have used $\tr \, Q_I = -\dot z_I$ and 
\beq
Q_I H_0 H_0^\dagger \Omega_g^{-1} \Omega_e^{-1} \big|_{z=z_I} = \lim_{z \rightarrow z_I} (z-z_I) \delta_t H_0 H_0^\dagger \Omega_g^{-1} \Omega_e^{-1} = 0.
\eeq
Note the strong similarity between Eq.~(\ref{eq:Leff1}) and Samols' formula \cite{Samols:1991ne} (although the analogy between the
definitions of our quantities $B_I$ and the corresponding ones of Samols is not complete, as it is made explicit in Section~\ref{sec:4}
Eq.~(\ref{eq:smallb})).

\subsection{Example: single vortex}\label{subsec:example}
As an example, let us consider the single vortex configuration. For the moduli matrix Eq.\,\eqref{eq:single}, the matrix $Q$ can be calculated as
\beq
Q = \lim_{z \rightarrow z_0} (z-z_0) \delta_t H_0 H_0^{-1} = 
\ba{cc} - \dot z_0 & 0 \\ 
-\dot{\vec \beta} & ~~\mathbf 0_{N-1} 
\ea.
\eeq
On the other hand, the matrix $B$ can be calculated by using the ansatz Eq.\,\eqref{eq:ansatz} as
\beq
B = - \lim_{z \rightarrow z_0} \left[ \frac{4}{m_g^2} \Omega_g \p \Omega_g^{-1} + \frac{4}{m_e^2} \Omega_e \p \Omega_e^{-1} \right] = -
\frac{4}{m_g^2} \lim_{z \rightarrow z_0} V \p V^{-1}.
\eeq 
Note that both functions $\psi_e$ and $\psi_g$ satisfy $\lim_{z \rightarrow z_0} \p \psi_e = \lim_{z \rightarrow z_0} \p \psi_g = 0$ (see Eq.\,\eqref{eq:expansion1} and Eq.\,\eqref{eq:expansion2}). From the explicit form of the matrix $V$ given in Eq.\,\eqref{eq:V}, we find that the matrix $B$ is given by
\beq
B = \frac{4}{m_g^2} \ba{cc} 0 & - \frac{\vec \beta^\dagger}{1+|\vec \beta|^2} \\ 0 & \mathbf 0_{N-1} \ea.
\eeq
Substituting $Q$ and $B$ into \eqref{eq:Leff1}, 
we obtain the following effective Lagrangian 
for a single non-Abelian vortex 
\beq
L_{\rm eff} = \pi v^2 |\dot z_0|^2 + \frac{4\pi}{g^2} \dot{\vec \beta}^\dagger \cdot \frac{ (1+|\vec \beta|^2) \mathbf 1 - \vec \beta \vec \beta^\dagger }{(1+|\vec \beta|^2)^2} \cdot \dot{\vec \beta}.
\eeq

\subsection{Coincident case}\label{subsec:coincident}
Next, let us consider the case of coincident vortices. If $\det H_0(z)$ has $k_I$-th order zero at $z=z_I$, the matrix $\delta_t H_0 H_0^{-1}$ has the following Laurent series expansion
\beq
\delta_t H_0 H_0^{-1} = \sum_{p=1}^{k_I} \frac{Q_p^I}{(z-z_I)^p} + \{ \mbox{regular at $z=z_I$} \}.
\eeq
On the other hand, the remaining part of the integrand is non-singular and can be expanded as 
\beq
\delta_t^\dagger \left( \frac{4}{m_g^2} \Omega_g \p \Omega_g^{-1} + \frac{4}{m_e^2} \Omega_e \p \Omega_e^{-1} \right) = \sum_{p,q=0}^\infty R_{p,q}^I (z-z_I)^p (\bar z-\bar z_I)^q.
\eeq
Then, the effective Lagrangian can be written in terms of the coefficients $Q_p$ and $R_{p,\,q}^I$ as
\beq
L_{\rm eff} = - \pi v^2 \sum_I \sum_{p=1}^{k_I} \tr ( Q_p^I R_{p-1,0}^I ).
\eeq

\section{Asymptotic metric for well-separated vortices}\label{sec:4}
In this section, we consider the asymptotic form of the 
metric on the moduli space for $k$ well-separated non-Abelian 
vortices by generalizing the results 
for Abelian vortices \cite{Manton:2002wb}.

\subsection{Asymptotic metric for Abelian vortices}
First, let us rederive  the effective Lagrangian for well-separated vortices in the $N=1$ (Abelian) theory \cite{Manton:2002wb}. Our
approach here has
essentially the same spirit of Section 2 in \cite{Manton:2002wb}, however our use of complex notation will make more transparent some
properties retained by the solutions of the linearized vortex equation, which are crucial for the derivation of the result.

In this
case, the moduli matrix $H_0(z)$ is a holomorphic polynomial of $z$ and can be written as
\beq
H_0(z) = \prod_{I=1}^k (z-z_I),
\eeq
where $k$ is the number of vortices and 
$z=z_I~(I=1,\cdots,k)$ are the positions of vortices. 
To calculate the asymptotic metric for well-separated vortices, it is convenient to define the function $\hat \psi$ by
\beq
\hat \psi \equiv \log \Omega_e - \log |H_0|^2 + \log v^2.
\eeq
Then, the master equation \eqref{eq:master} can be rewritten in terms of $\hat \psi$ as 
\beq
\frac{4}{m_e^2} \p \bar \p \hat \psi + e^{- \hat \psi} - 1 = - \frac{4 \pi}{m_e^2} \sum_{I=1}^k \delta^2(z-z_I),
\eeq
where boundary condition for large $|z|$ is given by $\hat \psi \rightarrow 0$.
Let us consider the linearized equation for the small fluctuation $\Delta \hat \psi$ around the background solution $\hat \psi$
\beq
\left( \frac{4}{m_e^2} \p \bar \p - e^{- \hat \psi} \right) \Delta \hat \psi = 0, \hs{10} z \not = z_I.
\label{eq:linear0}
\eeq
Let $\Delta_1 \hat \psi$ and $\Delta_2 \hat \psi$ be linearly independent solutions of the linearized equation. Then, the ``current" $(j_z,\,j_{\bar z})$ 
defined by\footnote{
Note that $j_z$ and $j_{\bar z}$ are not complex conjugate in general.
}
\beq
j_z \equiv \frac{4}{m_e^2} (\Delta_1 \hat \psi) \p (\Delta_2 \hat \psi) - (1 \leftrightarrow 2), \hs{10} j_{\bar z} \equiv \frac{4}{m_e^2} (\Delta_1 \hat \psi) \bar \p (\Delta_2 \hat \psi) - ( 1 \leftrightarrow 2),
\eeq
satisfies the following ``conservation law" except at the vortex positions 
\beq
\bar \p j_z + \p j_{\bar z} = 0, \hs{10} z \not = z_I.
\label{eq:CL}
\eeq
Therefore, the contour integrals
\beq
q_I \equiv \frac{1}{2\pi i} \oint_{C_I} ( dz j_z - d \bar z j_{\bar z} ), \hs{10} (I=1,\cdots,k)
\eeq
are invariant under continuous deformations of the contour $C_I$ surrounding $z=z_I$. This property of the invariants $q_I$ can
be used to
relate the local data in the neighborhood of each vortex to the asymptotic data.

As the first solution of the linearized equation, let us take 
the derivative of $\hat \psi$ with respect to the $I$-th moduli 
parameter $z_I$ (no sum over $I$ is implied) 
\beq
\Delta_1 \hat \psi ~=~ \delta_t \hat \psi ~\equiv~ \dot z_I \frac{\p \hat \psi}{\p z_I}.
\eeq
Note that $\delta_t \hat \psi$ satisfies the linearized equation \eqref{eq:linear0} except at $z=z_I$. As the second solution, we take the difference of the full $k$-vortex solution $\hat \psi$ and the single vortex solution $\hat \psi_I$ satisfying
\beq
\frac{4}{m_e^2} \p \bar \p \hat \psi_I + e^{- \hat \psi_I} + 1 = - \frac{4}{m_e^2} \delta^2(z-z_I).
\eeq
Note that the difference of the solutions 
\beq
\Delta_2 \hat \psi = \hat \psi - \hat \psi_I
\eeq
is an approximate solution of the linearized equation \eqref{eq:linear0} since both $\hat \psi$ and $\hat \psi_I$ satisfy the same equation except at $z=z_J~(J \not = I)$ and furthermore their difference is small if $|z-z_J| \gg m_e^{-1}$ for all $J \not = I$. 

Let us first calculate the contour integral by taking the zero radius limit of the circular contour $C_I$ surrounding $z=z_I$. Since the integral picks up the terms which behave like $1/(z-z_I)$ and $1/(\bar z - \bar z_I)$ in the zero radius limit, it can be evaluated by expanding the integrand around $z=z_I$
\beq
\hat \psi \, &=& - \log |z-z_I|^2 + a_I + b_I ( z-z_I ) + \bar b_I ( \bar z - \bar z_I ) + \frac{m_e^2}{4} |z-z_I|^2 + \cdots, \\
\hat \psi_I &=& - \log |z-z_I|^2 + \tilde a_I + \frac{m_e^2}{4} |z-z_I|^2 + \cdots,
\eeq
where the coefficients $b_I$ are related to 
$B_I = \frac{4}{m_e^2} \p \log \Omega_e |_{z=z_I}$ as\footnote{
Here the difference between our $B_I$ and Samols' $b_I$ is 
evident, however it is obvious that if one operates with 
$\delta_t^\dagger$, like in the 
final formula for the metric,  the outcome is the same.}
\beq
b_I \equiv \frac{m_e^2}{4} B_I - \sum_{J \not = I} \frac{1}{z_I-z_J}. \label{eq:smallb}
\eeq
Therefore, $\Delta_1 \hat \psi$ and $\Delta_2 \hat \psi$ behave around $z=z_I$ as
\beq
\Delta_1 \hat \psi &=& \frac{\dot z_I}{z-z_I} +\frac{\partial a_I}{\partial z_I} {\dot z_I}- b_I \dot z_I - \frac{m_e^2}{4} \dot z_I ( \bar
z - \bar z_I) + \cdots, \\
\Delta_2 \hat \psi &=& a_I 
- \tilde a_I + b_I ( z-z_I ) + \bar b_I ( \bar z - \bar z_I ) + \cdots.
\eeq
From this behavior of the functions $\Delta_1 \hat \psi$ and $\Delta_2 \hat \psi$, we obtain 
\beq
q_I ~=~ \frac{4}{m_e^2} 2 \dot z_I b_I ~=~ 
2 \dot z_I \left( B_I 
- {4\over m_e^2}\sum_{J \not = I} \frac{1}{z_I-z_J} \right) .
\eeq
On the other hand, the integral can also be evaluated along a large contour on which the solutions can be approximated by their asymptotic forms
\beq
\hat \psi &=& \sum_{J=1}^k c_e K_0(m_e |z-z_J|) + \cdots, \\
\hat \psi_I &=& c_e K_0(m_e |z-z_I|) + \cdots,
\eeq
Therefore, $\Delta_1 \hat \psi$ and $\Delta_2 \hat \psi$ have the following asymptotic forms
\beq
\Delta_1 \hat \psi &=& c_e \dot z_I \frac{\p}{\p z_I} K_0(m_e|z-z_I|) + \cdots , \\
\Delta_2 \hat \psi &=& \sum_{J \not = I} c_e K_0(m_e |z-z_J|) + \cdots,
\eeq
From these asymptotic forms of the functions $\Delta_1 \hat \psi$ and $\Delta_2 \hat \psi$, we obtain 
\beq
q_I &=& \frac{4 c_e^2}{m_e^2} \dot z_I \frac{\p}{\p z_I} \sum_{J \not = I}\int \left[ \left( \frac{dz}{2\pi i} K_0(m_e|z-z_I|) \p K_0(m_e|z-z_J|) - (I \leftrightarrow J) \right) + (c.c.) \right] \notag \\
&=& \frac{4c_e^2}{m_e^2} \dot z_I \frac{\p}{\p z_I} \sum_{J \not = I} K_0(m_e|z_I-z_J|).
\eeq
Comparing two expressions of $q_I$, 
we can relate the local data and the asymptotic data as
\beq
\dot z_I B_I = \sum_{J \not = I} \left( \frac{2c_e^2}{m_e^2} \dot z_I 
\frac{\p}{\p z_I} K_0(m_e|z_I-z_J|) 
+ {4\over m_e^2}\frac{\dot z_I}{z_I-z_J} \right).
\eeq
Since the constant $Q_I$ are given by
\beq
Q_I ~=~ \lim_{z \rightarrow z_I} (z-z_I) \delta_t H_0 H_0^{-1} ~=~ - \dot z_I,
\eeq
we obtain the effective Lagrangian from Eq.\,\eqref{eq:Leff1} as
\beq
L_{\rm eff} = \pi v^2 \left[ \sum_{I=1}^k |\dot z_I|^2 
- \frac{c_e^2}{4} \sum_{I \not = J} K_0( m_e |z_I - z_J| ) |\dot z_I - \dot z_J|^2 \right],
\eeq
where we have used (no sum over $I$ is implied) 
\beq
4 \frac{\p}{\p z_I} \frac{\p}{\p \bar z_I} K_0 (m_e|z_I-z_J|)  =
-4 \frac{\p}{\p z_I} \frac{\p}{\p \bar z_J} K_0 (m_e|z_I-z_J|)  =
 m_e^2 
 K_0 (m_e|z_I-z_J|) .
\eeq

\subsection{Asymptotic metric for non-Abelian vortices}
The most generic form moduli matrix for multi-vortex configuration is given by 
\beq
H_0 = \ba{cc} P(z) & 0 \\ \vec R(z) & \mathbf 1 \ea, \hs{5} P(z) = \prod_{I=1}^k (z-z_I), \hs{5} \vec R(z) = - \sum_{I=1}^k \vec \beta_I \prod_{J \not = I} \frac{z_{\phantom{I}}-z_J}{z_I-z_J}.
\eeq
There are $N$ complex moduli parameters for each vortex. One is the position moduli $z_I$ and the others are the orientational moduli $\vec \beta_I$. Note that the orientational vector for the vortex located at $z=z_I$ is 
\beq
\boldsymbol \phi \propto \ba{c} 1 \\ \vec \beta_I \ea, \hs{10} \bigg( \because H_0 \boldsymbol \phi \big|_{z=z_I} = 0 \bigg) .
\eeq
To calculate the asymptotic metric for well-separated vortices, it is convenient to redefine the matrices $\Omega_g$ and $\Omega_e$ by
\beq
\hat \Omega_g &\equiv& \phantom{v^2} |\det H_0|^{\frac{2}{N}} H_0^{-1} \Omega_g H_0^{\dagger-1}, \\
\hat \Omega_e &\equiv& v^2 |\det H_0|^{-\frac{2}{N}} \Omega_e .
\eeq
The matrices $\hat \Omega_e$ and $\hat \Omega_g$ satisfy 
\beq
\frac{4}{m_e^2} \bar \p ( \hat \Omega_e \p \hat \Omega_e^{-1} )+ \frac{4}{m_g^2} \bar \p ( \hat \Omega_g \p \hat \Omega_g^{-1} ) - \hat \Omega_g^{-1} \hat \Omega_e^{-1} + \mathbf 1_N = 0, \hs{10} z \not = z_I.
\label{eq:master-hat}
\eeq
The boundary conditions Eq.\,\eqref{eq:BD} can be translated into those for $\hat \Omega_e$ and $\hat \Omega_g$ as
\beq
\hat \Omega_e \rightarrow \mathbf 1_N, \hs{10} \hat \Omega_g \rightarrow \mathbf 1_N.
\eeq
For a given background solution $(\hat \Omega_e,\, \hat \Omega_g)$, let us consider the following linearized equation for small fluctuations $(\Delta \hat \Omega_e,\,\Delta \hat \Omega_g)$
\beq
\frac{4}{m_e^2} \bar \p \left[ \Delta (\hat \Omega_e \p \hat \Omega_e^{-1} ) \right] + \frac{4}{m_g^2} \bar \p \left[ \Delta (\hat \Omega_g \p \hat \Omega_g^{-1}) \right] - \hat \Omega_g^{-1} \hat \Omega_e^{-1} ( \hat \Omega_g \Delta \hat \Omega_g^{-1} + \hat \Omega_e \Delta \hat \Omega_e^{-1} ) = 0, \hs{5} z \not = z_I.
\label{eq:linearizedEq}
\eeq
As in the case of the Abelian vortices, we can define the 
``conserved current''$(j_z,j_{\bar z})$ from the solutions of the linearized equation \eqref{eq:linearizedEq} as 
\beq
j_z &=& \tr \left[ \frac{4}{m_e^2} (\hat \Omega_e \Delta_1 \hat \Omega_e^{-1}) \Delta_2 ( \hat \Omega_e \p \hat \Omega_e^{-1} ) + \frac{4}{m_g^2} ( \hat \Omega_g \Delta_1 \hat \Omega_g^{-1} ) \Delta_2 (\hat \Omega_g \p \hat \Omega_g^{-1} ) \right] - (1 \leftrightarrow 2), \\
j_{\bar z} &=& \tr \left[ \frac{4}{m_e^2} ( \Delta_1 \hat \Omega_e^{-1} \hat \Omega_e ) \Delta_2 ( \bar \p \hat \Omega_e^{-1} \hat \Omega_e ) + \frac{4}{m_g^2} ( \Delta_1 \hat \Omega_g^{-1} \hat \Omega_g ) \Delta_2 ( \bar \p \hat \Omega_g^{-1} \hat \Omega_g ) \right] - (1 \leftrightarrow 2). 
\eeq
As the solution of the linearized equation, we take the derivatives of $\hat \Omega_e$ and $\hat \Omega_g$ with respect to the moduli parameters
\beq
\Delta_1 \hat \Omega_e = \delta_t \hat \Omega_e, \hs{10} \Delta_1 \hat \Omega_g = \delta_t \hat \Omega_g,
\eeq
and the difference of the full solution $(\hat \Omega_e,\,\hat \Omega_g)$ and the single vortex solution $(\hat \Omega_{eI},\,\hat \Omega_{gI})$
\beq
\Delta_2 \hat \Omega_e = \hat \Omega_e - \hat \Omega_{eI}, \hs{10} \Delta_2 \hat \Omega_g = \hat \Omega_g - \hat \Omega_{gI}. 
\eeq
From the fact that the current $(j_z,j_{\bar z})$ satisfies the conservation law $\bar \p j_z + \p j_{\bar z} =0$, the contour integrals
\beq
q_I = \frac{1}{2\pi i} \oint_{C_I} \left( dz \, j_z - d \bar z \, j_{\bar z} \right), \hs{5} (I=1,2,\cdots,k)
\label{eq:contour}
\eeq
are invariant under the continuous deformation of the contour $C_I$ surrounding $z=z_I$. By expanding the integrand around $z=z_I$ and picking up the terms which behave like $1/(z-z_I)$ and $1/(\bar z - \bar z_I)$, the contour integral $q_I$ can be evaluated in the zero-radius limit as
\beq
q_I &=& 2 \, \tr \Big[ Q_I B_I \Big] - 2 \,\tr \left[ Q_I^{\rm single} B_I^{\rm single} \right] + f(z_J, \beta_J) \notag \\
&=& 2 \, \tr \left[ Q_I B_I \right] - \frac{8}{m_g^2} 
\frac{\vec \beta_I^\dagger\cdot \dot{\vec \beta}_I }{1+|\vec \beta_I|^2} + f(z_J, \beta_J), 
\label{eq:q_small}
\eeq
where $f(z_J, \beta_J)$ is a holomorphic function of the moduli parameters. 
Next, let us evaluate the integral along a large contour on which the
solutions can be approximated by their asymptotic forms 
Eq.(\ref{eq:asym1}) and Eq.(\ref{eq:asym2})
\beq
\hat \Omega_e &=& \mathbf 1_N + c_e \sum_{I=1}^k K_0(m_e|z-z_I|) \mathbf 1_N + \cdots, \\
\hat \Omega_g &=& \mathbf 1_N + c_g \sum_{I=1}^k K_0(m_g|z-z_I|) T_{\beta_I} + \cdots. 
\eeq
By using these asymptotic forms and the formula
\beq
K_0 ( m |z_I - z_J | ) = \int_{C_I} \frac{dz}{2\pi i} \Big[ K_0( m |z-z_I| ) \p K_0 ( m |z-z_J| ) - ( I \leftrightarrow J ) \Big] + (c.c.),
\eeq
we obtain 
\beq
q_I = \delta_t \left[ \frac{4Nc_e^2}{m_e^2} \sum_{J \not = I} K_0(m_e|z_I-z_J|) + \frac{4Nc_g^2}{m_g^2} \sum_{J \not = I} \Theta_{IJ} K_0 (m_g|z_I-z_J|) \right],
\label{eq:q_large}
\eeq
where $\Theta_{IJ}$ is defined by
\beq
\Theta_{IJ} ~\equiv~ \frac{1}{N} \tr ( T_{\beta_I} T_{\beta_J} ) ~=~ N \frac{|1+\vec \beta_I \cdot \vec \beta_J^\dagger|^2}{(1+|\vec \beta_I|^2)(1+|\vec \beta_J|^2)} - 1 .
\eeq
Comparing Eq.\,\eqref{eq:q_small} and Eq.\,\eqref{eq:q_large}, we find that 
\beq
\tr \left[ Q_I \delta_t^\dagger B_I \right] &=& - \delta_t^\dagger \delta_t \left[ \frac{2Nc_e^2}{m_e^2} \sum_{J \not = I} K_0(m_e|z_I-z_J|) + \frac{2Nc_g^2}{m_g^2} \sum_{J \not = I} \Theta_{IJ} K_0 (m_g|z_I-z_J|) \right] \notag \\
&{}& +\delta_t^\dagger \delta_t \left[ \frac{4}{m_g^2} \log (1+|\vec \beta_I|^2) \right].
\eeq
Then, we obtain the asymptotic effective Lagrangian form Eq.\,\eqref{eq:Leff1} 
\beq
L_{\rm eff} ~=~ \pi v^2 \left( \sum_{I=1}^k |\dot z|^2 + \tr \left[ Q_I \delta_t^\dagger B_I \right] \right) ~=~ \delta_t^\dagger \delta_t K, 
\label{eq:Leff}
\eeq
where the K\"ahler potential is given by
\beq
K &=& \sum_{I=1}^k \left[ \pi v^2 |z_I|^2 + \frac{4\pi}{g^2} \log ( 1 + |\vec \beta_I|^2 ) \right] - \sum_{I \not = J} \left[\frac{2 \pi v^2 Nc_e^2}{m_e^2} K_0(m_e|z_I-z_J|) \right] \notag \\
&-& \sum_{I \not = J} \left[\frac{2 \pi v^2 N c_g^2}{m_g^2} \left( N \frac{|1+\vec \beta_I \cdot \vec \beta_J^\dagger|^2}{(1+|\vec \beta_I|^2)(1+|\vec \beta_J|^2)} - 1 \right) K_0(m_g|z_I-z_J|) \right]. 
\label{eq:K}
\eeq
The last two terms describe 
leading interactions between different vortices.
In homogeneous coordinates ($N$-vector) $\vec{\phi}_I$ 
for each ${\mathbb C}P^{N-1}$, after a K\"ahler transformation
the K\"ahler potential can be rewritten as 
\beq
K &=& \sum_{I=1}^k \left[ \pi v^2 |z_I|^2 + \frac{4\pi}{g^2} \log |\vec \phi_I|^2  \right] - \sum_{I \not = J} \left[\frac{2 \pi v^2 Nc_e^2}{m_e^2} K_0(m_e|z_I-z_J|) \right] \notag \\
&-& \sum_{I \not = J} \left[\frac{2 \pi v^2 N c_g^2}{m_g^2} \left( N \frac{|\boldsymbol \phi_I \cdot \boldsymbol \phi_J^\dagger|^2}{|\boldsymbol \phi_I|^2|\boldsymbol \phi_J|^2} - 1 \right) K_0(m_g|z_I-z_J|) \right]. 
\label{eq:Khom}
\eeq
For the case $m_e \ge 2m_g$, the second terms of 
Eq.(\ref{eq:K}) and Eq.(\ref{eq:Khom}) are incorrect 
since we have used Eq.(\ref{eq:asym1}) 
as an asymptotic behavior of $\psi_e$. 
They should be replaced with terms of order $e^{-2m_g|z_I-z_J|}$. 
However, the dominant contribution to the interaction  comes from the last term, which is of order $e^{-m_g|z-z_0|}$.
Therefore we can say that the above results are correct even in the case of 
$m_e\ge 2m_g$ if we neglect subleading contributions to the interaction.

\section{The point source formalism} \label{sec:5} 
It has been shown that the interaction between well-separated Abelian vortices can be identified with that between composites of point-like scalar source and magnetic dipole \cite{Manton:2002wb,Speight:1996px}. 
In \cite{Eto:2008mf} some of us used the point particle approximation 
for vortex-strings stretched between domain walls 
and found a good agreement with the direct calculation.

In this section, we show that a non-Abelian vortex looks like a point particle in the following linear field theory
\beq
\mathcal L &=& + \frac{1}{2} ( \p_\mu \Phi^0 \p^\mu \Phi^0 - m_e^2 \Phi^0 \Phi^0) - \frac{1}{4e^2} \left( f_{\mu \nu} f^{\mu \nu} - 2 m_e^2 w_\mu w^\mu \right) \notag \\
&{}& + \frac{1}{2} ( \p_\mu \Phi^a \p^\mu \Phi^a - m_g^2 \Phi^a \Phi^a ) - \frac{1}{4g^2} \left( F_{\mu \nu}^a F^{a \mu \nu } - 2 m_g^2 W_\mu^a W^{a \mu} \right) \notag \\
&{}& + \kappa^0 \Phi^0 + \kappa^a \Phi^a - \frac{1}{e^2} j^{0 \mu} w_\mu - \frac{1}{g^2} j^{a \mu} W_\mu^a. \phantom{\frac{1}{2}}
\eeq
Here, $\Phi^0$ and $\Phi^a~(a=1,\cdots,N^2-1)$ are scalar fields, $w_\mu$ and $W_\mu^a~(a=1,\cdots,N^2-1)$ are the massive vector fields and $\kappa^0$, $\kappa^a$, $j_\mu^0$, $j_\mu^a$ are point-like sources corresponding to a non-Abelian vortex. 
Let us consider the following scalar source $\kappa \equiv \kappa^0 t^0 + \kappa^a t^a$
\beq
\kappa = - \pi v \left[ 1 - \frac{|\dot z_0|^2}{2} + \frac{1}{m_g^2} \left( \nabla_t \dot \beta^i \frac{\p}{\p \beta^i} + \overline{\nabla_t \dot \beta^i} \frac{\p}{\p \bar \beta^i} \right) \right] (c_e \mathbf 1_N + c_g T_\beta) \delta^2(z-z_0), \label{eq:kappa} 
\eeq
where $T_\beta$ is defined in Eq.~\eqref{eq:tbeta} 
and $\nabla_t \dot \beta^i$ is defined by
\beq
\nabla_t \dot \beta^i \equiv \left( \p_t - \frac{2 \bar \beta^j \dot \beta^j}{1+|\beta^j|^2} \right) \dot \beta^i.
\eeq
For the vector source $j_\mu \equiv j_\mu^0 t^0 + j_\mu^a t^a$, we assume the following form
\beq
j^\mu = \p_\nu X^{\mu \nu}, \label{eq:jmu}
\eeq
where $X^{\mu \nu}$ is an anti-symmetric tensor. Note that this form of the vector current implies the conservation law $\partial_\mu  j^\mu=0$. For the anti-symmetric tensor $X^{\mu \nu}$, let us consider the following form 
\beq
X^{\mu \nu} = - \pi \left[ \epsilon^{\mu \nu \rho} J_\rho + \frac{1}{m_g^2} ( \p^\mu J^\nu - \p^\nu J^\mu ) i \Big( \dot \beta^i \frac{\p}{\p \beta^i} - \dot{\bar \beta}^i \frac{\p}{\p \bar \beta^i} \Big) \right] ( c_e \mathbf 1_N + c_g T_\beta ). 
\eeq
where $J^\mu$ is the current of a point particle 
\beq
J^\mu \equiv \dot x^\mu_0 \delta^2(z-z_0). 
\eeq
Now we will see that a non-Abelian vortex, viewed from distance, looks like the above point source at least up to second order in
the time-derivative. We can check this by comparing the field configuration for the point source and the asymptotic  fields for a
non-Abelian vortex derived in Appendix \ref{appendix:DE}. 
The equations of motion for the scalar fields $\Phi^0$, $\Phi^a$ and the vector fields $w_\mu$, $W_\mu$ are 
\beq
\kappa^0 &=& (\p_\mu \p^\mu + m_e^2 ) \Phi^0, \hs{10} 
\kappa^a ~=~ (\p_\mu \p^\mu + m_g^2 ) \Phi^a, \label{eq:linear1} \\ 
j_\nu^0 &=& (\p_\mu \p^\mu + m_e^2 ) w_\nu, \hs{10}  
j_\nu^a ~=~ (\p_\mu \p^\mu + m_g^2 ) W^{a}_{\nu}, \label{eq:linear2}
\eeq
where we have assumed that the vector fields 
satisfy the Lorenz gauge condition 
\beq
\p_\mu w^\mu = \p_\mu W^{a\mu} = 0.
\eeq
Operating $\partial^\nu$  
on both sides of \eqref{eq:linear2}, 
we see that this gauge condition implies current conservation. The
solution of the equations \eqref{eq:linear1} and \eqref{eq:linear2} can be obtained by using the Green's function expanded in terms of the
time-derivative 
\beq
\Phi^0 &=& \frac{1}{2\pi} \int d^2 z' \left[ K_0(m_e|z-z'|) + \frac{1}{m_e^2} \Upsilon(m_e|z-z'|) \p_t^2 + \mathcal O(\p_t^4) \right] \kappa^0 (t,z'),
\label{eq:phi} \\
\Phi^a &=& \frac{1}{2\pi} \int d^2 z' \left[ K_0(m_g|z-z'|) + \frac{1}{m_g^2} \Upsilon(m_g|z-z'|) \p_t^2 + \mathcal O(\p_t^4) \right] \kappa^a(t,z'),
\label{eq:Phi} \\
w_\mu &=& \frac{1}{2\pi} \int d^2 z' \left[ K_0(m_e|z-z'|) + \frac{1}{m_e^2} \Upsilon(m_e|z-z'|) \p_t^2 + \mathcal O(\p_t^4) \right] j_\mu^0(t,z'),
\label{eq:w} \\
W_\mu^a &=& \frac{1}{2\pi} \int d^2 z' \left[ K_0(m_g|z-z'|) + \frac{1}{m_g^2} \Upsilon(m_g|z-z'|) \p_t^2 + \mathcal O(\p_t^4) \right] j_\mu^a(t,z').
\label{eq:W} 
\eeq
where 
\begin{equation}
\Upsilon(s)=\frac{1}{2} s K_0'(s)
. 
\label{eq:def_upsilon}
\end{equation}
It is convenient to expand the sources in terms of the time-derivative 
\beq
\kappa = \kappa^{(0)} + \kappa^{(1)} + \kappa^{(2)} + \mathcal O(\p_t^3) , \hs{10} j_\mu = j_\mu^{(0)} + j_\mu^{(1)} + j_\mu^{(2)} + \mathcal O(\p_t^3).
\eeq
Correspondingly, the scalar field $\Phi \equiv \Phi^0 t^0 + \Phi^a t^a$ and the vector field $W_\mu = w_\mu t^0 + W_\mu^a t^a$ are expanded as
\beq
\Phi = \Phi^{(0)} + \Phi^{(1)} + \Phi^{(2)} + \mathcal O(\p_t^3) ,
\hs{10} 
W_\mu = W_\mu^{(0)} + W_\mu^{(1)} + W_\mu^{(2)} + \mathcal O(\p_t^3) .
\eeq
Let us check that the field configuration for the point source and that for a non-Abelian vortex order by order.  Note that odd (even) order
equations of motion for $\Phi^{(n)}$, $w_{\bar z}^{(n)}$ and $W_{\bar z}^{(n)}$ ($w_0^{(n)}$ and $W_0^{(n)}$) are trivial due to 
time-reversal symmetry.
First, we consider the leading terms $\Phi^{(0)}$ and $W_\mu^{(0)}$ corresponding to the static configuration. The leading order terms of the sources are given by
\beq
\kappa^{(0)} = - \pi v (c_e \mathbf 1_N + c_g T_\beta) \delta^2(z-z_0), 
\hs{10}
j_{\bar z}^{(0)} = - \pi i ( c_e \mathbf 1_N + c_g T_{\beta} ) \bar \p \delta^2 (z-z_0).
\eeq
Note that $j_0$ has no term without time-derivative. Evaluating \eqref{eq:phi}, \eqref{eq:Phi}, \eqref{eq:w} and \eqref{eq:W}, we obtain the following solutions for the static source 
\beq
\Phi^{(0)} = - \frac{v}{2} \Psi, 
\hs{10}
W_{\bar z}^{(0)} = - \frac{i}{2} \bar \p \Psi. 
\label{eq:0th}
\eeq
where the function $\Psi$ is given by
\beq
\Psi ~\equiv~ c_e \mathbf 1_N K_0(m_e|z-z_0|) + c_g T_{\beta} K_0(m_e|z-z_0|).
\eeq
These solutions agree with the asymptotic forms of the static single vortex solution \eqref{eq:staticH} and \eqref{eq:staticW} in a singular gauge. 
Next, let us consider the first order contribution which are contained only in $j_0$
\beq
j_0^{(1)} = - \pi i \left[ \dot z_0 \frac{\p}{\p z_0} - \dot{\bar z}_0 \frac{\p}{\p \bar z_0} + \frac{4}{m_g^2} \frac{\p}{\p z_0} \frac{\p}{\p \bar z_0} \left( \dot \beta^i \frac{\p}{\p \beta^i} - \dot{\bar \beta}^i \frac{\p}{\p \bar \beta^i} \right) \right] ( c_e \mathbf 1_N + c_g T_{\beta} ) \delta^2(z-z_0).
\label{eq:1st}
\eeq
By using \eqref{eq:w}, \eqref{eq:W} and $(4 \p \bar \p - m^2 ) K_0(m|z-z_0|) = 0~(z \not = z_0)$, we obtain 
\beq
W_0^{(1)} &=& \frac{i}{2} \left[ \dot z_0 \frac{\p}{\p z_0} - \dot{\bar z}_0 \frac{\p}{\p \bar z_0} + \dot \beta^i \frac{\p}{\p \beta^i} - \dot{\bar \beta}^i \frac{\p}{\p \bar \beta^i} \right] \Psi. \label{eq:sol_w0}
\eeq
The second order terms in the scalar source $\kappa$ and the current $j_{\bar z}$ are given by
\beq
\kappa^{(2)} &=& \pi v \left[ \frac{|\dot z_0|^2}{2} - \frac{1}{m_g^2} \left( \nabla_t \dot \beta^i \frac{\p}{\p \beta^i} + \overline{\nabla_t \dot \beta^i} \frac{\p}{\p \bar \beta^i} \right) \right] \delta^2(z-z_0) (c_e \mathbf 1_N + c_g T_\beta), \\
j_{\bar z}^{(2)} 
&=& - \frac{\pi i}{2} \left( \ddot z_0 + \dot z_0 \frac{\p}{\p z_0} + \dot{\bar z}_0 \frac{\p}{\p \bar z_0} \right) \delta^2(z-z_0) ( c_e \mathbf 1_N + c_g T_{\beta} ) \notag \\
&{}& - \frac{2\pi}{m_g^2} i \left( \dot \beta^i \frac{\p}{\p \beta^i} - \dot{\bar \beta}^i \frac{\p}{\p \bar \beta^i} \right) \dot z_0 \frac{\p}{\p z_0} \frac{\p}{\p \bar z_0} \delta^2(z-z_0) ( c_e \mathbf 1_N + c_g T_{\beta} ) \notag \\
&{}& - \frac{\pi i}{m_g^2} \left( \nabla_t \dot \beta^i \frac{\p}{\p \beta^i} - \nabla_t \dot{\bar \beta}^i \frac{\p}{\p \bar \beta^i} \right) \frac{\p}{\p \bar z_0} \delta^2(z-z_0) ( c_e \mathbf 1_N + c_g T_{\beta} ).
\eeq
By evaluating the integral \eqref{eq:phi}, \eqref{eq:Phi}, \eqref{eq:w} and \eqref{eq:W}, we obtain the following solutions 
\beq
\Phi^{(2)} &=& \phantom{-} \frac{1}{8} v \left[ \ddot z_0 (\bar z - \bar z_0) + \ddot{\bar z}_0 ( z - z_0 ) \right] \Psi \notag \\
&{}& - \frac{1}{8} v \left[ \dot z_0 (\bar z - \bar z_0) + \dot{\bar z}_0 (z - z_0) \right] ( \dot z_0 \p + \dot{\bar z}_0 \bar \p) \Psi \notag \\
&{}& + \frac{1}{4} v \left[ \dot z_0 (\bar z - \bar z_0) + \dot{\bar z}_0 (z - z_0) \right] \left( \dot \beta^i \frac{\p}{\p \beta^i} + \dot{\bar \beta}^i \frac{\p}{\p \bar \beta^i} \right) \Psi \notag \\
&{}& - \frac{1}{2m_g^2} v \left( \nabla_t \dot \beta^i \frac{\p}{\p \beta^i} + \overline{\nabla_t \dot \beta^i} \frac{\p}{\p \bar \beta^i} \right) \Psi \notag \\
&{}& - \frac{1}{2m_g^2} v c_g \Upsilon(m_g|z-z_0|) \p_t^2 T_{\beta}, \label{eq:sol2_phi}
\eeq
\beq
W_{\bar z}^{(2)} &=& \phantom{+} \frac{i}{8} \left[ \ddot z_0 (\bar z - \bar z_0) + \ddot{\bar z}_0 (z - z_0) \right] \bar \p \Psi \notag \\
&{}& - \frac{i}{8} \left[ \dot z_0 (\bar z - \bar z_0) + \dot{\bar z}_0 (z - z_0) \right] \left( \dot z_0 \p + \dot{\bar z}_0 \bar \p \right) \bar \p \Psi \notag \\
&{}& - \frac{i}{8} \left[ \ddot z_0 - \dot z_0 ( \dot z_0 \p - \dot{\bar z}_0 \bar \p) \right] \Psi \notag \\
&{}& + \frac{i}{4} \left[ \dot z_0 (\bar z - \bar z_0) + \dot{\bar z}_0 (z - z_0) \right] \left( \dot \beta^i \frac{\p}{\p \beta^i} + \dot{\bar \beta}^i \frac{\p}{\p \bar \beta^i} \right) \bar \p \Psi \notag \\
&{}& - \frac{i}{m_g^2} \dot z_0 \left( \dot \beta^i \frac{\p}{\p \beta^i} - \dot{\bar \beta}^i \frac{\p}{\p \bar \beta^i} \right) \p \bar \p \Psi \notag \\
&{}& + \frac{i}{2m_g^2} \left( \nabla_t \dot \beta^i \frac{\p}{\p \beta^i} - \overline{\nabla_t \dot \beta^i} \frac{\p}{\p \bar \beta^i} \right) \bar \p \Psi \notag \\
&{}& - \frac{i}{2m_g^2} c_g \bar \p \Upsilon(m_g|z-z_J|) \p_t^2 T_{\beta}, \label{eq:sol2_W}
\eeq
As shown in the Appendix \ref{appendix:DE}, the solutions \eqref{eq:sol_w0}, \eqref{eq:sol2_phi} and \eqref{eq:sol2_W} agree with those obtained by using the method of derivative expansion with a moving vortex background. 

Let us rederive the effective Lagrangian for the non-Abelian vortices Eq.\,\eqref{eq:Leff} from the point particle formalism. The effective Lagrangian for the point particles are given by
\beq
L_{\rm eff} = \sum_{I=1}^k \left( \pi v |\dot z_I|^2 + \frac{4\pi}{g^2} \dot \beta^i \dot{\bar \beta}^j \frac{\p^2}{\p \beta^i \bar \beta^j} \log ( 1 + |\beta^i|^2 ) \right) + \sum_{I > J} L_{(I,J)}.
\eeq
Here, $L_{(I,J)}$ is the interaction Lagrangian between $I$-th and $J$-th particles
\beq
L_{(I,J)} = \int d^2 x \left( \kappa^0_I \Phi_J^0 + \kappa^a_I \Phi^a_J - \frac{1}{e^2} j^{0 \mu}_I w_{\mu J} - \frac{1}{g^2} j^{a \mu}_I W_{\mu J}^a \right),
\eeq
$\kappa^0_I$, $\kappa^a_I$, $j_{\mu I}^0$ and $j_{\mu I}^a$ are the point-like source with $z_0 = z_I$, $\beta^i = \beta^i_I$ and $\Phi_J^0$, $\Phi_J^a$, $w_{\mu J}$ and $W_{\mu J}$ are the fields induced by the point-like source with $z_0 = z_J$, $\beta^i = \beta^i_J$. First, let us consider static interaction 
\beq
L_{(I,J)}^{(0)} &=& \frac{1}{2} \int d^2 x \Bigg[ \kappa_I^{0(0)} \Phi_J^{0(0)} + \frac{4}{e^2} j_{zI}^{0(0)} w_{\bar z J}^{(0)} + \kappa_I^{a(0)} \Phi_J^{a(0)} + \frac{4}{g^2} j_{zI}^{a(0)} W_{\bar z J}^{a(0)} + (c.c.) \Bigg].
\eeq
Since the static sources $\kappa_I^{(0)}$ and $j_{z I}^{(0)}$ are given by
\beq
\kappa_I^{(0)} = - \pi v ( c_e \mathbf 1_N + c_g T_{\beta_I} ) \delta^2( z - z_I ), \hs{10}
j_{z I}^{(0)} = \pi i ( c_e \mathbf 1_N + c_g T_{\beta_I} ) \p \delta^2(z-z_I), \label{eq:0-th}
\eeq 
the static interaction Lagrangian reduces to
\beq
L_{(I,J)}^{(0)} = - \pi v \, {\rm tr} \left[ \left( \Phi_J^{0(0)} t^0 + \frac{4v}{m_e^2} i \p w_{\bar z J}^{(0)} t^0 + \Phi_J^{a(0)} t^a + \frac{4v}{m_g^2} i \p W_{\bar z J}^{a(0)} t^a \right) ( c_e \mathbf 1_N + c_g T_{\beta_I} ) \right]_{z=z_I}.
\eeq
We can show by using Eq.\,\eqref{eq:0th} and $(4 \p \bar \p - m^2 ) K_0(m|z-z_0|) = 0$ that
\beq
\left( \Phi_J^{0(0)} t^0 + \frac{4v}{m_e^2} i \p W_{\bar z J}^{0(0)} t^0 \right) = \left( \Phi_J^{a(0)} t^a + \frac{4v}{m_g^2} i \p W_{\bar z J}^{a(0)} t^a \right) = 0.
\eeq
Therefore, the static interaction vanishes as expected from the BPS property. 
The leading order terms in the interaction Lagrangian are second order in the time-derivative 
\beq
L_{(I,J)}^{(2)} &=& \int d^2 x \Bigg[ {\rm Re} \left( \kappa_I^{0(0)} \Phi_J^{0(2)} + \frac{4}{e^2} j_{z I}^{0(0)} W_{\bar z J}^{0(2)} + \kappa_I^{a(0)} \Phi_J^{a(2)} + \frac{4}{g^2} j_{zI}^{a(0)} W_{\bar z J}^{a(2)} \right) + (I \leftrightarrow J ) \notag \\
&{}& \hs{10} - \p_t \Phi_I^{0(0)} \p_t \Phi_J^{0(0)} - \p_t \Phi_I^{a(0)} \p_t \Phi_J^{a(0)} - \frac{4}{e^2} \p_t W_{zI}^{0(0)} \p_t W_{\bar zJ}^{0(0)} - \frac{4}{g^2} \p_t W_{zI}^{a(0)} \p_t W_{\bar zJ}^{a(0)} \notag \\
&{}& \hs{10} - \frac{1}{e^2} j_{0I}^{0(1)} W_{0J}^{0(1)} - \frac{1}{g^2} j_{0I}^{a(1)} W_{0J}^{a(1)} \Bigg].
\eeq
Here, we have ignored total time derivatives and used 
\beq
\int d^2 x \, \kappa_I^{0(2)} \Phi_J^{0(0)} &=& \int d^2 x \left[ \p_t \p_t \Phi_I^{0(0)} \Phi_J^{0(0)} + \Phi_I^{0(2)} \kappa_J^{0(0)} \right], 
\eeq
and similar identities for the other fields. To calculate the interaction Lagrangian $L_{(I,J)}^{(2)}$, it is convenient to decompose it as 
\beq
\mathcal A_e &\equiv& \int d^2 x \, {\rm Re} \left( \kappa_I^{0(0)} \Phi_J^{0(2)} + \frac{4}{e^2} j_{zI}^{0(0)} W_{\bar z J}^{0(2)} \right), \\
\mathcal A_g &\equiv& \int d^2 x \, {\rm Re} \left( \kappa_I^{a(0)} \Phi_J^{a(2)} + \frac{4}{g^2} j_{zI}^{a(0)} W_{\bar z J}^{a(2)} \right), \\
\mathcal B_1 &\equiv& \int d^2 x \left( \p_t \Phi_I^{0(0)} \p_t \Phi_J^{0(0)} + \p_t \Phi_I^{a(0)} \p_t \Phi_J^{a(0)} \right), \\
\mathcal B_2 &\equiv& \int d^2 x \left( \frac{4}{e^2} \p_t W_{zI}^{0(0)} \p_t W_{\bar zJ}^{0(0)} + \frac{4}{g^2} \p_t W_{zI}^{a(0)} \p_t W_{\bar zJ}^{a(0)} \right), \\
\mathcal C &\equiv& \int d^2 x \left( \frac{1}{e^2} j_{0I}^{0(1)} W_{0J}^{0(1)} + \frac{1}{e^2} j_{0I}^{a(1)} W_{0J}^{a(1)} \right).
\eeq
By using the explicit form of the static sources Eq.\,\eqref{eq:0-th} and the solutions \eqref{eq:sol2_phi} and \eqref{eq:sol2_W}, we obtain
\beq
\mathcal A_e + \mathcal A_g &=& -2 \pi v \, {\rm Re} \left[ c_e \tr \left( \Phi_J^{0(2)} t^0 + \frac{4v}{m_e^2} i \p W_{\bar z J}^{0(2)} t^0 \right) + c_g \tr \left( \Phi_J^{a(2)} t^a + \frac{4v}{m_e^2} i \p W_{\bar z J}^{a(2)} t^a \right) \right]_{z=z_I} \notag \\
&=& - \delta_J \delta_J^\dagger \left[ \frac{2 \pi c_e^2}{e^2} N K_0(m_e|z_I-z_J|) + \frac{2 \pi c_e^2}{g^2} \tr [ T_{\beta_I} T_{\beta_J} ] K_0(m_g|z_I-z_J|) \right],
\eeq
where we have defined the differential operators $\delta_J,\,\delta_J^\dagger$ by
\beq
\delta_J \equiv \dot z_J \frac{\p}{\p z_J} + \dot \beta_J^i \frac{\p}{\p \beta_J^i}, \hs{10} \delta_J^\dagger \equiv \dot{\bar z}_J \frac{\p}{\p \bar z_J} + \dot{\bar \beta}_J^i \frac{\p}{\p \bar \beta_J^i}.
\eeq
By using the identity
\beq
\int d^2 x \, K_0(m |z-z_I|) K_0(m |z-z_J|) 
= - \frac{2 \pi}{m^2} \Upsilon(m |z_I-z_J|) 
\eeq
with $\Upsilon$ in Eq.(\ref{eq:def_upsilon}), 
we can calculate $\mathcal B_1$ and $\mathcal B_2$ as
\beq
\mathcal B_1 
&=& ( \delta_I + \delta_I^\dagger ) ( \delta_J + \delta_J^\dagger ) \left[ - \frac{\pi c_e^2}{e^2} N \Upsilon(m_e|z_I-z_J|) - \frac{\pi c_g^2}{g^2} \tr \left[ T_{\beta_I} T_{\beta_J} \right] \Upsilon(m_g|z_I-z_J|) \right], \\
\mathcal B_2 
&=& ( \delta_I + \delta_I^\dagger ) ( \delta_J + \delta_J^\dagger ) \notag \\
&{}& \hs{8} \times \frac{\p}{\p z_I} \frac{\p}{\p \bar z_J} \left[ - \frac{4 \pi c_e^2}{m_e^2 e^2} N \Upsilon(m_e|z_I-z_J|) - \frac{4\pi c_g^2}{m_g^2 g^2} \tr \left[ T_{\beta_I} T_{\beta_J} \right] \Upsilon(m_g|z_I-z_J|) \right].
\eeq
Therefore, we find that
\beq
\mathcal B_1 + \mathcal B_2 = ( \delta_I + \delta_I^\dagger ) ( \delta_J + \delta_J^\dagger ) \left[ \frac{\pi c_e^2}{e^2} N K_0(m_e|z_I-z_J|) + \frac{\pi c_g^2}{g^2} \tr \left[ T_{\beta_I} T_{\beta_J} \right] K_0(m_g|z_I-z_J|) \right], 
\eeq
where we have used $( - 4 \p \bar \p + m^2 ) \Upsilon(m|z-z_0|) = - m^2 K_0(m|z-z_0|)$. From \eqref{eq:1st} and \eqref{eq:sol_w0}, we obtain
\beq
\mathcal C = - ( \delta_I - \delta_I^\dagger ) ( \delta_J - \delta_J^\dagger ) \Big[ \frac{\pi c_e^2}{e^2} N K_0(m_e|z_I-z_J|) + \frac{\pi c_g^2}{g^2} \tr \left[ T_{\beta_I} T_{\beta_J} \right] K_0(m_g|z_I-z_J|) \Big]. 
\eeq
Therefore, the interaction Lagrangian is given by
\beq
L_{(I,J)} &=& \left[ \mathcal A_e + \mathcal A_g + (I \leftrightarrow J) \right] - \left[ \mathcal B_1 + \mathcal B_2 + \mathcal C \right] \\
&=& - (\delta_I + \delta_J)(\delta_I^\dagger + \delta_J^\dagger) \left[ \frac{2\pi c_e^2}{e^2} N K_0(m_e|z_I-z_J|) + \frac{2\pi c_g^2}{g^2} \tr \left[ T_{\beta_I} T_{\beta_J} \right] K_0(m_g|z_I-z_J|) \right] . \notag
\eeq
This interaction Lagrangian agrees with that in the effective Lagrangian Eq.\,\eqref{eq:Leff}. 

\section{Conclusions and Discussions}

By exploiting the local nature of non-Abelian vortices 
for $N_{\rm C}=N_{\rm F}$, we have extended the Samols' formula  
of for the metric on the moduli space 
to the non-Abelian $U(N)$ case. 
This fact has enabled us to construct the explicit metric 
(\ref{eq:Leff}) and its K\"ahler potential (\ref{eq:K})
on the moduli space of well-separated non-Abelian vortices. 
We have also {derived the metric using 
an appropriate} point-particle approximation for non-Abelian vortices. 

In this paper we have studied {\it local} vortices 
in $U(N)$ gauge theory with the 
same number of fundamental Higgs fields as the number $N$ of colors.
When the flavor number is greater than the color number, 
the vortices are {instead} called {\it semi-local} 
\cite{Vachaspati:1991dz}.
A typical property of semi-local vortices is that 
they have a size modulus which is in general 
non-normalizable. 
In the case of a semi-local non-Abelian vortex,
its orientational moduli are also non-normalizable \cite{Shifman:2006kd}
unless its size modulus vanishes\cite{Eto:2007yv}. 
The (non-)normalizability of zero modes 
was completely classified in \cite{Eto:2007yv} 
for arbitrary number of vortices 
with arbitrary moduli. 
Since wave functions for non-normalizable moduli are 
divergent in infinite space, 
we can have a metric only for the normalizable 
moduli. 
The most direct way to extend the present work 
to the case of semi-local non-Abelian vortices 
would be the calculation of such metric through the generalization of Samols' formula. 
However, the metric of well-separated semi-local vortices should be 
well-approximated by that of lumps \cite{Ward:1985ij}, 
since the asymptotic behavior in this case is 
efficiently described by lumps rather than local vortices. 
Therefore one should be able to work out the metric without 
the generalization of  Samols' formula \cite{Leese:1992fn}. 

In order to address the metric problem in different gauge groups, such as $SO$ and $Sp$, one must take into account the
observation that
 non-Abelian vortices 
are generically semi-local. This means that our result can  be immediately 
generalized only to the particular sector of their moduli space where 
vortices happen to be local, while for the full space a supplementary work of the type mentioned in the previous paragraphs will be needed.
Non-Abelian vortices with gauge group $G \times U(1)$ 
with arbitrary simple group $G$ were constructed in \cite{Eto:2008yi}. 
It was found for instance that 
the moduli spaces of single local vortices in 
$G=SO(2N),USp(2N)$ gauge theories are 
Hermitian symmetric spaces $SO(2N)/U(N)$ and $USp(2N)/U(N)$, 
respectively.
The moduli space of non-Abelian vortices in 
$SO$ and $USp$ gauge theories were further studied 
in  \cite{Ferretti:2007rp,Eto:2008qw,Eto:2009bg,Gudnason:2010jq}.

Other possible directions are:  
\begin{itemize}
 \item [i)] to obtain the generalization of the Samols' formula 
to {geometries} other than ${\mathbb C}$, 
for instance, a cylinder \cite{Eto:2006mz}, 
a torus \cite{Eto:2007aw}, 
Riemann surfaces \cite{Baptista:2008ex} 
or a hyperbolic space 
\cite{Witten:1976ck,Krusch:2009tn,Manton:2010wu}, 
\item[ii)] the inclusion of a Chern-Simons term \cite{Collie:2008mx},
\item[iii)] the generalization to the non-BPS case where a potential term 
will be induced on the moduli space \cite{Auzzi:2007wj}.
\end{itemize}

\section*{Acknowledgments}

We would like to thank Minoru Eto for discussion in the early stage of this work.
G.~M. acknowledges support from ``A.~Della Riccia'' Foundation and Japan Society for Promotion of Science. 
The work of M.N.~and of N.S~are supported in part by 
Grant-in-Aid for Scientific 
Research No.~20740141 (M.N.), No.~21540279 (N.S.) and 
No.~21244036 (N.S.) from the Ministry 
of Education, Culture, Sports, Science and Technology-Japan.

\appendix 
\section{Derivative expansion}\label{appendix:DE}
In this section, we will see that the solutions \eqref{eq:sol2_phi} and \eqref{eq:sol2_W} agree with the asymptotic fields for the vortex configuration by using the derivative expansion.  
It is convenient to use a singular gauge by performing the following singular gauge transformations $\mathcal U_e \in U(1)$ and $\mathcal U_g \in SU(N)$
\beq
\mathcal U_e = \left( \frac{\bar z - \bar z_0}{|z-z_0|} \right)^{\frac{1}{N} \mathbf 1_N }, \hs{10} \mathcal U_g = U^\dagger \left( \frac{\bar z - \bar z_0}{|z-z_0|} \right)^{\frac{1}{N} T} U.
\eeq
Then, the solutions for a single vortex in a non-singular 
gauge Eq.\,\eqref{eq:single_S} become
\beq
S_e = v^{-1} (z-z_0)^{\frac{1}{N} \mathbf 1_N} e^{\frac{1}{2} \hat \psi_e \mathbf 1_N}, \hs{10} S_g = V \, (z-z_0)^{\frac{1}{N} T} e^{\frac{1}{2} \hat \psi_g T} \, U, 
\eeq
where $\hat \psi_e$ and $\hat \psi_g$ are defined by
\beq
\hat \psi_e &\equiv& \psi_e - \frac{1}{N} \log |z-z_0|^2 ~\approx~ c_e K_0(m_e|z-z_0|) + \cdots, \label{eq:asy_psi_e} \\
\hat \psi_g &\equiv& \psi_g - \frac{1}{N} \log |z-z_0|^2 ~\approx~ c_g K_0(m_g|z-z_0|) + \cdots. \label{eq:asy_psi_g}
\eeq
Note that $S_e S_g$ is single-valued whereas each of $S_e$ and
$S_g$ is multi-valued.
Correspondingly, the solutions for the Higgs fields and the gauge fields \eqref{eq:single_H}, \eqref{eq:single_w}, \eqref{eq:single_W} become
\beq
H ~=~ v \, e^{-\frac{1}{2} (\hat \psi_e \mathbf 1_N + \hat \psi_g T_\beta)}, \hs{10}
w_{\bar z} t^0 + W_{\bar z}^a t^a ~=~ - \frac{i}{2} \bar \p ( \hat \psi_e \mathbf 1_N + \hat \psi_g T_\beta ), 
\eeq
where $T_\beta$ is the matrix defined in Eq.\,\eqref{eq:tbeta}. 
We focus on the asymptotic form of the single vortex solution in the singular gauge
\beq
H - v \mathbf 1_N ~~ &=& - \frac{1}{2} v c_e K_0 ( m_e |z-z_0| ) \mathbf 1_N - \frac{1}{2} v c_g K_0 ( m_g |z-z_0| ) T_\beta + \cdots , \label{eq:staticH} \\
w_{\bar z} t^0 + W_{\bar z}^a t^a &=& - \frac{i}{2} c_e \bar \p K_0 ( m_e |z-z_0| ) \mathbf 1_N - \frac{i}{2} c_g \bar \p K_0 ( m_g |z-z_0|) T_\beta + \cdots, \label{eq:staticW}
\eeq
Identifying the scalar fields $\Phi \equiv \Phi^0 t^0 + \Phi^a t^a$ with
$H - v \mathbf 1_N$ in the singular gauge, we find that the terms without time-derivative in \eqref{eq:0th} agree with the static solutions \eqref{eq:staticH} and \eqref{eq:staticW}. To check that the terms with time-derivative are also correct, let us consider a time-dependent background configuration by promoting the moduli parameters to time dependent dynamical valuables. 
The time-dependence induces the fluctuations around the background configuration, so we expand the fields with respect to the number of the time-derivatives 
\beq
H(t) &=& H^{\rm BPS} (z_0(t), \beta^i(t)) + H^{(1)}(t) + H^{(2)}(t) + \cdots,
\eeq
and similarly for the vector fields $w_\mu$ and $W_\mu^a$. 
As a background solution, let us consider a single BPS vortex configuration satisfying the following BPS equations
\beq
\D_{\bar z} H ~=~ 0, \hs{10} \frac{4}{e^2} i f_{z \bar z} t^0 + \frac{4}{g^2} i F_{z \bar z}^a t^a + H H^\dagger - v^2 \mathbf 1_N ~=~ 0.
\eeq
If the position moduli $z_0$ and the orientation $\beta^i$ are constant, this background configuration satisfies the equations of motion
\beq
0 &=& \D_\mu \D^\mu H + e^2 ( H_A^\dagger t^0 H_A - \xi ) t^0 H + g^2 ( H_A^\dagger t^a H_A ) t^a H , \label{eq:eom1} \\
0 &=& \D^\mu \left( \frac{2}{e^2} f_{\mu \nu} t^0 + \frac{2}{g^2} F_{\mu \nu}^a t^a \right) + i \Big[ H (\D_\nu H)^\dagger - (\D_\nu H) H^\dagger \Big]. \label{eq:eom2}
\eeq
If we give weak time dependences to the moduli parameters
\beq
z_0 \rightarrow z_0(t), \hs{10} \beta^i \rightarrow \beta^i(t), 
\eeq
Eq.\,\eqref{eq:eom1} and Eq.\,\eqref{eq:eom2} become equations of motion for the fluctuations, which can be solve order-by-order. Note that odd (even) order equations of motion for $H^{(n)}$, $w_{\bar z}^{(n)}$ and $W_{\bar z}^{(n)}$ ($w_0^{(n)}$ and $W_0^{(n)}$) are trivial due to the time-reversal symmetry. The first order term in the gauge fields $w_0$ and $W_0^a$ are determined from the Gauss' law equation
\beq
0 &=& - \D_z \left( \frac{4}{e^2} f_{\bar z 0} t^0 + \frac{4}{g^2} F_{\bar z 0}^a t^a \right) - i (\D_0 H) H^\dagger + (h.c.).
\eeq
This equation can be solved as
\beq
w_0^{(1)} t^0 ~=~ i ( \delta_t S_e^\dagger S_e^{\dagger-1} - S_e^{-1} \delta_t^\dagger S_e ), \hs{10}
W_0^{a (1)} t^a ~=~ i ( \delta_t S_g^\dagger S_g^{\dagger-1} - S_g^{-1} \delta_t^\dagger S_g ). \label{eq:W0}
\eeq
By using the asymptotic forms \eqref{eq:asy_psi_e} and \eqref{eq:asy_psi_g}, we can check that Eq.\,\eqref{eq:sol_w0} agrees with the asymptotic forms of $w_0$ and $W_0$.

The asymptotic behavior of the second order fluctuations are determined from the linearized equations of motion
\beq
- \frac{1}{4} \D_0 \D_0 H ~~~~~~~ &=& i \D_z \mathcal H + \frac{1}{2} ( \mathcal W_e + \mathcal W_e^\dagger + \mathcal W_g + \mathcal W_g^\dagger ) H, \label{eq:linear_a} \\
-\D_0 \left[ \frac{1}{e^2} f_{0 \bar z} t^0 + \frac{1}{g^2} F_{0 \bar z}^a t^a \right] &=& i \D_{\bar z} \left[ \frac{2}{e^2} (\mathcal W_e + \mathcal W_e^\dagger) + \frac{2}{g^2} (\mathcal W_g + \mathcal W_g^\dagger) \right] - \mathcal H H^\dagger. \label{eq:linear_b}
\eeq
Here, we have defined $\mathcal H$, $\mathcal W_e$ and $\mathcal W_g$ by
\beq
\mathcal H~ &\equiv& i \D_{\bar z} H^{(2)} - (w_{\bar z}^{(2)} t^0 + W_{\bar z}^{a (2)} t^a ) H , \label{eq:H} \\ 
\mathcal W_e &\equiv& i \D_z (w_{\bar z}^{(2)} t^0) + \frac{e^2}{2} \tr ( H^{(2)} H^\dagger t^0 ) t^0, \label{eq:W_e} \\
\mathcal W_g &\equiv& i \D_z ( W_{\bar z}^{a (2)} t^a ) + \frac{g^2}{2} \tr ( H^{(2)} H^\dagger t^a) t^a . \label{eq:W_g}
\eeq
Note that the linearized equations of motion have the physical zero mode associated with the moduli parameters and the gauge zero modes corresponding to the gauge transformation acting on the fluctuations. 
The gauge transformation acts of the fluctuation fields as
\beq
H^{(2)} \rightarrow H^{(2)} + i A^{(2)} H, \hs{10} w_{\bar z}^{(2)} t^0 + W_{\bar z}^{a (2)} t^a \rightarrow w_{\bar z}^{(2)} t^0 + W_{\bar z}^{a (2)} t^a - \D_{\bar z} A^{(2)}, 
\label{eq:gauge_trans}
\eeq
where $A^{(2)}$ is an arbitrary hermitian matrix which 
is of second order in the time-derivative. To solve the equations \eqref{eq:linear_a} and \eqref{eq:linear_b}, it is convenient to take the following gauge for the fluctuations
\beq
\mathcal W_e = \mathcal W_e^\dagger, \hs{10} \mathcal W_g = \mathcal W_g^\dagger.
\eeq
Then, the equations \eqref{eq:linear_a} and \eqref{eq:linear_b} can be solved with respect to $\mathcal H$, $\mathcal W_e$ and $\mathcal W_g$ as
\beq
\mathcal W_e &=& - \frac{1}{4} S_e^{-1} S_g^{-1} \left[ \dot \phi^i \dot{\bar \phi}^j \frac{\p}{\p \bar \phi^j} \left(\Omega_e \frac{\p}{\p \phi^i} \Omega_e^{-1} \right) \right] S_e S_g, \\
\mathcal W_g &=& - \frac{1}{4} S_e^{-1} S_g^{-1} \left[ \dot \phi^i \dot{\bar \phi}^j \frac{\p}{\p \bar \phi^j} \left(\Omega_g \frac{\p}{\p \phi^i} \Omega_g^{-1} \right) \right] S_e S_g,
\eeq
\beq
\mathcal H = S_e^\dagger S_g^\dagger \left[ \left( \ddot \phi^i \frac{\p}{\p \phi^i} + \dot \phi^i \dot \phi^j  \frac{\p}{\p \phi^i} \frac{\p}{\p \phi^j} \right) \left( \frac{1}{e^2} \bar \p \Omega_e^{-1} \Omega_e + \frac{1}{g^2} \bar \p \Omega_g^{-1} \Omega_g \right) \right] H_0^{\dagger-1}.
\eeq
The fluctuations $H^{(2)}$, $w_{\bar z}^{(2)}$ and $W_{\bar z}^{a(2)}$ are determined from \eqref{eq:H}, \eqref{eq:W_e} and \eqref{eq:W_g}. We can assume the following form
\beq
H^{(2)} \hs{8} &=& G - i \D_z \left( \frac{4}{e^2} X^0 t^0 + \frac{4}{g^2} X^a t^a \right) H^{\dagger-1}, \\
w_{\bar z}^{(2)} t^0 + W_{\bar z}^{a(2)} t^a &=& i (\D_{\bar z} G) H^{-1} + X^0 t^0 + X^a t^a, 
\eeq
where $G$, $X^0$ and $X^a$ are the fields satisfying
\beq
\mathcal H \hs{7} &=& \D_{\bar z} \left[ \D_z \left( \frac{4}{e^2} X^0 t^0 + \frac{4}{g^2} X^a t^a \right) H^{\dagger-1} \right] - (X^0 t^0 + X^a t^a) H, \label{eq:X} \\
\mathcal W_e + \mathcal W_g &=& - \D_z ( \D_{\bar z} G H^{-1} ) + \frac{e^2}{2} \tr (G H^\dagger t^0) t^0 + \frac{g^2}{2} \tr (G H^\dagger t^a) t^a. \label{eq:G}
\eeq
If we focus on their asymptotic forms, the background fields can be replaced with their vacuum expectation values. Therefore, the equations \eqref{eq:G} and \eqref{eq:X} reduces to
\beq
\ddot \phi^i \frac{\p}{\p \phi^i} \Psi + \dot \phi^i \dot \phi^j \frac{\p}{\p \phi^i} \frac{\p}{\p \phi^j} \Psi &=& 
- \bar \p \p \left( \frac{4}{m_e^2} X^0 t^0 + \frac{4}{m_g^2} X^a t^a \right) + (X^0 t^0 + X^a t^a), \\
v \, \dot \phi^i \dot{\bar \phi}^j \frac{\p}{\p \phi^i} \frac{\p}{\p \bar \phi^j} \Psi &=& - 4 \p \bar \p G + 2 m_e^2 \tr (G t^0) t^0 + 2 m_g^2 \tr (G t^a) t^a. 
\eeq
These equations can be solved as
\beq
G \hs{8} &=& - \frac{v}{4} (\bar z - \bar z_0) \dot z_0 \dot{\bar z}_0 \bar \p \Psi - \frac{v}{m_g^2} (z-z_0) \dot \beta^i \dot{\bar \beta}^j \frac{\p^2}{\p \beta^i \p \bar \beta^j} \p \Psi \notag \\
&{}& + \frac{v}{4} \left[ (\bar z - \bar z_0) \dot z_0 \dot{\bar \beta}^i \frac{\p}{\p \bar \beta^i} + (z-z_0) \dot{\bar z}_0 \dot \beta^i \frac{\p}{\p \beta^i} \right] \Psi, \\
X^0 t^0 + X^a t^a &=& + \frac{i}{4} (\bar z - \bar z_0) \left( \ddot z_0 - \dot z_0 \dot z_0 \p \right) \bar \p \Psi \notag \\
&{}& + \frac{i}{2} (\bar z - \bar z_0) \dot z_0 \dot \beta^i \frac{\p}{\p \beta^i} \bar \p \Psi - \frac{i}{4} (z-z_0) \nabla_t \dot \beta^i \frac{\p}{\p \beta^i} \Psi.
\eeq
Then, we obtain a solution for the second order fluctuations as
\beq
H^{(2)} 
&=& \phantom{+} \frac{v}{4} (\bar z - \bar z_0) \left[ \ddot z_0 - \dot z_0 ( \dot z_0 \p + \dot{\bar z}_0 \bar \p) \right] \Psi \notag \\ 
&{}& + \frac{v}{4} \left[ (\bar z - \bar z_0) \dot z_0 \dot{\bar \beta}^i \frac{\p}{\p \bar \beta^i} + (z-z_0) \dot{\bar z}_0 \dot \beta^i \frac{\p}{\p \beta^i} \right] \Psi \notag \\
&{}& + \frac{v}{2} (\bar z - \bar z_0) \dot z_0 \dot \beta^i \frac{\p}{\p \beta^i} \Psi - \frac{v}{m_g^2} (z-z_0) \dot \beta^i \dot{\bar \beta}^j \frac{\p^2}{\p \beta^i \p \bar \beta^j} \p \Psi \notag \\
&{}& - \frac{v}{m_g^2} \nabla_t \dot \beta^i \frac{\p}{\p \beta^i} \p  \left[ (z-z_0) \Psi \right] + \{ \mbox{zero modes} \}, \phantom{\frac{1}{2}}
\eeq
\beq
w_{\bar z}^{(2)} t^0 + W_{\bar z}^{a (2)} t^a 
&=& \phantom{+} \frac{i}{4} \left[ \ddot z_0 - \dot z_0 \left( \dot z_0 \p + \dot{\bar z}_0 \bar \p \right) \right] \left[ (\bar z - \bar z_0) \bar \p \Psi \right] \notag \\
&{}& + \frac{i}{4} \left[ \dot z_0 \dot{\bar \beta}^i \frac{\p}{\p \bar \beta^i} + \dot{\bar z}_0 \dot \beta^i \frac{\p}{\p \beta^i} \right] \bar \p \left[ (\bar z - \bar z_0) \Psi \right] \notag \\
&{}& + \frac{i}{2} (\bar z - \bar z_0) \dot z_0 \dot \beta^i \frac{\p}{\p \beta^i} \bar \p \Psi - \frac{i}{4} (z-z_0) \left( \dot \beta^i \dot{\bar \beta}^j \frac{\p^2}{\p \beta^i \p \bar \beta^j} + \nabla_t \dot \beta^i \frac{\p}{\p \beta^i} \right) \Psi \notag \\
&{}& + \{ \mbox{zero modes} \}. \phantom{\frac{1}{2}}
\eeq
We can show that these solutions are related to \eqref{eq:sol2_phi} and \eqref{eq:sol2_W} by the gauge transformation \eqref{eq:gauge_trans} with 
\beq
A^{(2)} &=& \bigg[ \frac{i}{8} (\bar z - \bar z_0) \left[ \ddot z_0 - \dot z_0 ( \dot z_0 \p + \dot{\bar z}_0 \bar \p) \right] \Psi + \frac{i}{4} (\bar z - \bar z_0) \dot z_0 \dot \beta^i \frac{\p}{\p \beta^i} \Psi \notag \\
&{}& - \frac{ic_g}{2m_g^2} \nabla_t \dot \beta^i \frac{\p}{\p \beta^i} \p \left[ (z-z_0) \Psi \right] \bigg] + (h.c.).
\eeq

\end{document}